\documentclass[11pt]{article} 
\usepackage{fullpage,citesort,amssymb,epsfig,amsmath,psfrag} 
\usepackage[normal,small]{caption} 
\newcommand{\beq}{\begin{equation}} 
\newcommand{\eeq}{\end{equation}} 
\newcommand{\bea}{\begin{eqnarray}} 
\newcommand{\eea}{\end{eqnarray}}

\def\Tr{{\rm Tr}} 
\def\to{\rightarrow}

\def\ie{{\it i.e.\ }} 
\def\m@th{\mathsurround=0pt } 
\def\leftrightarrowfill{$\m@th \mathord\leftarrow \mkern-6mu \cleaders\hbox{$\mkern-2mu \mathord- \mkern-2mu$}\hfill 
 \mkern-6mu \mathord\rightarrow$} 
\def\overleftrightarrow#1{\vbox{\ialign{##\crcr 
     \leftrightarrowfill\crcr\noalign{\kern-1pt\nointerlineskip} 
     $\hfil\displaystyle{#1}\hfil$\crcr}}}

\begin{document} 
 
\renewcommand{\thefootnote}{\fnsymbol{footnote}} 
\begin{titlepage} 
\begin{flushright} 
UFIFT-HEP-02-12\\ 
hep-ph/0203232 
\end{flushright} 
 
\vskip 2.5cm 
 
\begin{center} 
\begin{Large} 
{\bf One Loop Calculations in Gauge Theories\\ Regulated on an $x^+$-$p^+$ 
Lattice\footnote{This work was supported in part by the Department 
of Energy under Grant No. DE-FG02-97ER-41029.} 
} 
\end{Large} 
 
\vskip 2.cm 
 
{\large  
Skuli Gudmundsson\footnote{E-mail  address: {\tt skulig@phys.ufl.edu}}  
and Charles B. Thorn\footnote{E-mail  address: {\tt thorn@phys.ufl.edu}}} 
 
\vskip 0.5cm

{\it Institute for Fundamental Theory\\ 
Department of Physics, University of Florida, 
Gainesville, FL 32611} 
 
(\today) 
 
\vskip 1.0cm 
\end{center} 
 
\begin{abstract} 
In earlier work, the planar diagrams of $SU(N_c)$ gauge 
theory have been regulated on the light-cone by a scheme  
involving both discrete 
$p^+$ and $\tau=ix^+$. The transverse 
coordinates remain continuous, but even so all diagrams are 
rendered finite by this procedure.  In this scheme quartic interactions 
are represented as two cubics mediated by short lived 
fictitious particles whose detailed behavior could be adjusted   
to retain properties of the continuum theory, at least at one loop.  
%serve as counter-terms 
Here we use this setup to  
calculate the one loop three gauge boson triangle diagram, 
and so complete the calculation of diagrams renormalizing the 
coupling to one loop.  
In particular, we find that the cubic vertex is correctly renormalized 
once the couplings to the fictitious particles are chosen 
to keep the gauge bosons massless. 
\end{abstract} 
% 
%\pacs{} 
\vfill 
\end{titlepage} 
% 
%\narrowtext
\begin{section}{Introduction} 
 
The large $N_c$ limit of $SU(N_c)$ gauge theory, introduced 
long ago by 't Hooft \cite{thooftlargen}, remains unsolved 
even though the limit singles out the planar Feynman 
diagrams of perturbation theory. Although a dramatic 
simplification, the sum of all planar diagrams still 
represents a rich enough dynamics to frustrate many 
imaginative approaches to their solution  
\cite{largenattempts}. But even if an exact analytic solution is 
out of reach, this should not mean that the nonperturbative physics 
of the limit is hopelessly intractable. Indeed, in the 
face of the even richer dynamics of gauge theory at finite $N_c$, much 
insight has been gleaned by numerically studying lattice 
gauge theory  
\cite{wilsonconfine,lattice2001,Campostrini:2000zn} 
on a finite lattice  
as a useful nonperturbative model of the 
continuum gauge theory. Moreover, since the lattice could in 
principle be taken ever larger in size, lattice gauge theory  
can provide a concrete definition of continuum gauge theory. 
 
Of course, one approach to the large $N_c$ limit would be 
simply to study lattice gauge theory for ever larger $N_c$ 
\cite{Lucini:2001ej}.  
This is certainly a well-posed and 
interesting formulation of the problem. 
However, it is 
somewhat removed from the appealing and 
intuitive idea \cite{nielsenfishnet}  
that the large planar diagrams known as fishnets  
could, in a confining theory, 
provide a model of a QCD string ``world-sheet''. This idea 
is now even more compelling because of the conjectured 
equivalence of certain supersymmetric large $N_c$ 
gauge theories to supergravity or superstring theories 
\cite{maldacena}. It would be nice to have an explicit 
discretized model representation of the sum of planar diagrams, 
which, if analytic methods fail, can at least be analyzed 
numerically. 
Accordingly, Bering, Rozowsky, and Thorn (BRT) \cite{beringrt} 
reconsidered and refined earlier attempts \cite{thornfishnet,gilesmt,browergt}  
to construct a lattice model 
of the sum of planar diagrams by working in an infinite 
momentum frame (\ie on a light-front) and discretizing 
$\tau=ix^+$, imaginary light-cone time, and $p^+$, the kinematic 
light-cone momentum ($p^-$ is the light-cone Hamiltonian).  
In this approach there is no need to discretize the transverse 
coordinate space or transverse momentum space, since the 
discretization of $\tau$ and $p^+$, with the exclusion 
of the zero values of these variables,  cuts off all ultraviolet and 
infra-red divergences. This approach has been further developed 
in Ref.~\cite{bardakcit}. 
 
In addition to setting up the basic formulation of the 
$x^+,p^+$ lattice, the authors of \cite{beringrt} also 
calculated the one-loop gluon self-energy  
diagram as a check of the faithfulness of the 
lattice as a regulator of divergences. Although agreement 
with an earlier continuum calculation was obtained for 
this diagram, one still needs to calculate the 
one loop three gluon triangle diagram to check whether the 
regulation procedure gives the correct charge renormalization. 
We expand on the work of \cite{beringrt} here by doing additional one loop 
calculations in the pure $SU(N_c)$ gauge theory 
using the BRT discretization. In particular, we want to check 
that this scheme does not disturb the light-cone gauge asymptotic freedom 
calculations done with a simple momentum cutoff in  
Ref.~\cite{thornfreedom,perrylc}. We 
obtain expressions for the triangle diagram to one loop order and 
show that color charge is indeed correctly renormalized as in \cite{perrylc}. 
Further, we extract all of the divergences  arising from the 
infra-red region of small $p^+$. Although in individual diagrams 
there are double logarithms, we show that in the complete 
sum only single logs remain. 
 
The Feynman rules used in \cite{beringrt} are unusual 
in several respects. First, the basic propagators 
are given in the mixed $ix^+,p^+,{\bf p}$ representation, 
with $ix^+=ka$ and $p^+=lm$, with $k,l=1,2,3,\ldots$.  
The transverse polarization of the gluon is given 
in the complex basis $\wedge=(1+i2)/\sqrt2$, $\vee=(1-i2)/\sqrt2$, 
represented graphically by an arrow attached to the  
transverse gluon line. 
Next, since we focus on the planar diagrams of the $N_c\to\infty$ 
limit, there is no need for the double line notation, and all 
$N_c$ dependence can be absorbed in the coupling constant  
$g\equiv g_s\sqrt{N_c}$. 
The resulting vertices for this restricted context accordingly 
depend on the cyclic ordering of the lines entering the 
vertex. Finally, all the quartic vertices, including 
those induced by integrating out $A_+$ in $A_-=0$ 
gauge, are represented as the concatenation of  
two cubic vertices, with fictitious particles mediating 
the quartic interaction. The one mediating the induced 
quartic interaction (the instantaneous``Coulomb'' exchange)  
can be thought of as a remnant of the 
$A_+$ field, whereas that mediating the $\Tr[A_k,A_l]^2$ 
vertex can be thought of as a remnant of $F_{kl}$ in 
the first order form of the action. These fictitious 
particles would not propagate in the continuous time 
$a\to0$ limit, but with $a$ finite are allowed to 
propagate a limited number of time steps. This is implemented 
by including a $k$-dependent factor $f_k$ or $h_k$ in the propagator 
for each fictitious particle. These must satisfy 
$\sum_k f_k=\sum_k h_k=1$ and must fall off rapidly 
with $k$, in order that the correct tree amplitudes 
be correctly produced. The Lorentz invariance of perturbation theory 
puts further constraints on their $k$-dependence. 
Indeed, the flexibility offered by tuning these 
coefficients may even obviate the need for 
further explicit counter-terms to guarantee Lorentz invariance.  
One constraint has already been put on this 
behavior in \cite{beringrt} by requiring the gauge 
boson to remain massless to one loop. We shall find that 
with no further constraints, color charge is 
correctly renormalized. 
 
The final set of Feynman rules for the transverse gauge 
bosons and the fictitious scalars in the 
context of planar diagrams is summarized in Fig.~\ref{NewRules} 
\begin{figure}[ht] 
\begin{center} 
\begin{tabular}{|c|c||c|c|} 
\hline 
&&&\\[-.3cm] 
$ 
\begin{array}[c]{c} 
\psfig{file=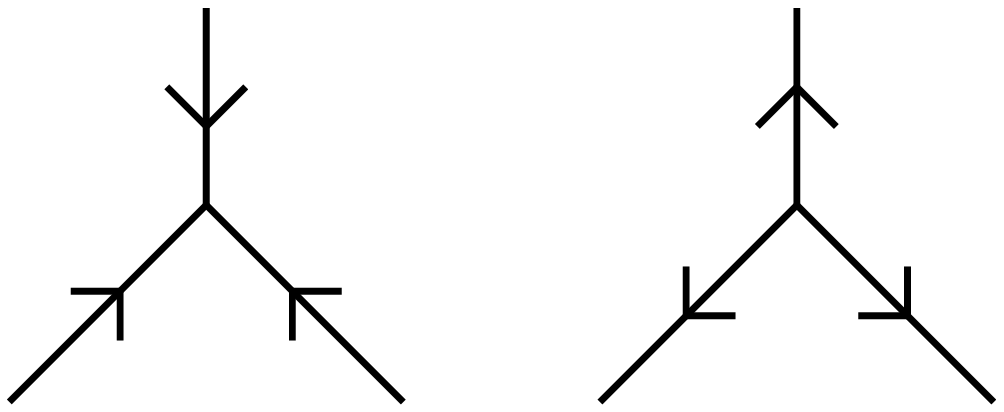,height=0.5in} 
\end{array} 
$ 
&  
$0$ 
& 
$ 
\begin{array}[c]{c} 
\psfig{file=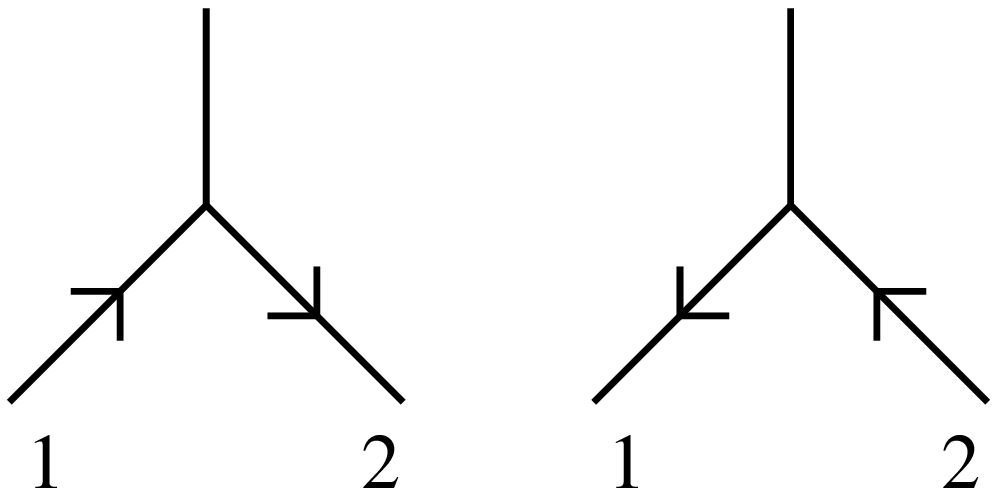,height=0.6in} 
\end{array} 
$ 
&  
$ {g\over T_0}(M_2-M_1)$ \\ 
\hline 
&&&\\[-.3cm] 
$ 
\begin{array}[c]{c} 
\psfig{file=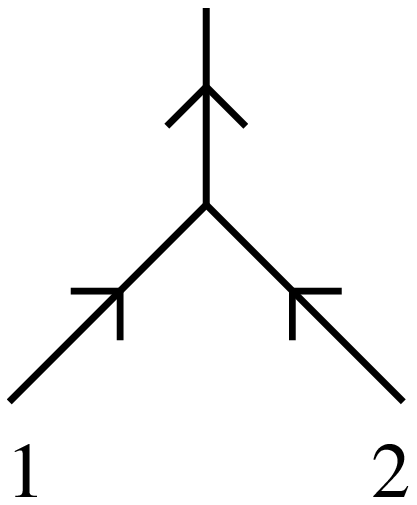,height=0.6in} 
\end{array} 
$ 
&  
$ -2{g\over T_0} \left({M_1+M_2\over M_1M_2}\right) 
(M_1Q^\wedge_2-M_2Q^\wedge_1) $ 
& 
$ 
\begin{array}[c]{c} 
\psfig{file=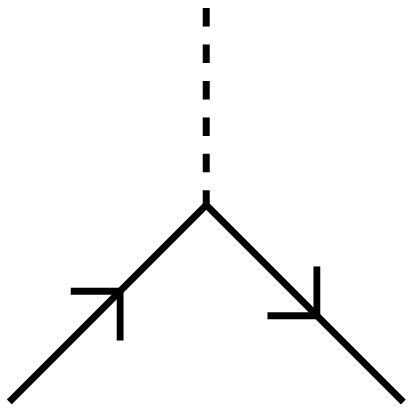,height=0.6in} 
\end{array} 
$ 
&  
$+{g\over T_0}$ \\ 
\hline 
&&&\\[-.3cm] 
$ 
\begin{array}[c]{c} 
\psfig{file=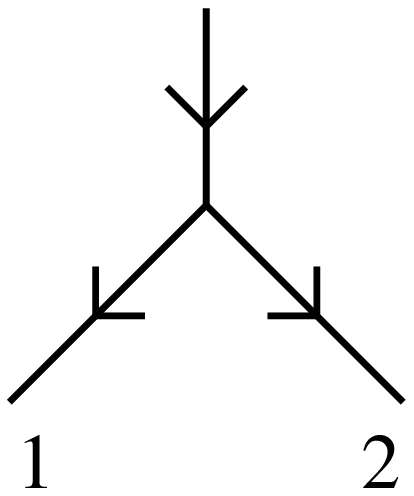,height=0.6in} 
\end{array} 
$ 
&  
$ -2{g\over T_0} \left({M_1+M_2\over M_1M_2}\right) 
(M_1Q^\vee_2-M_2Q^\vee_1)$ 
& 
$ 
\begin{array}[c]{c} 
\psfig{file=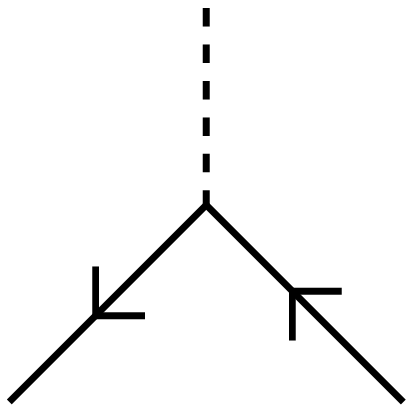,height=0.6in} 
\end{array} 
$ 
&  
$ -{g\over T_0}$ \\ 
\hline 
&\multicolumn{3}{c|}{}\\[-.2cm] 
$ 
\begin{array}[c]{c} 
\psfig{file=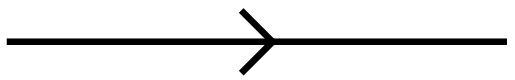,width=0.6in} 
\end{array} 
$ 
& 
\multicolumn{3}{c|}{ 
${1 \over 2M} e^{-k{\bf Q}^2/2MT_0}$ 
} \\[.3cm]  
\hline 
&\multicolumn{3}{c|}{}\\[-.2cm] 
$ 
\begin{array}[c]{c} 
\psfig{file=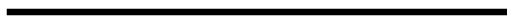 ,width=0.6in} 
\end{array} 
$ 
& 
\multicolumn{3}{c|}{ 
$-f_k {T_0\over M^{2}}e^{-k{\bf Q}^2/2MT_0}$ 
} \\[.3cm]  
\hline 
&\multicolumn{3}{c|}{}\\[-.2cm] 
$ 
\begin{array}[c]{c} 
\psfig{file=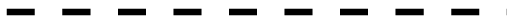,width=0.6in} 
\end{array} 
$ 
& 
\multicolumn{3}{c|}{ 
$-h_k T_0 e^{-k{\bf Q}^2/2MT_0}$ 
} \\[.3cm]  
\hline 
\end{tabular} 
\end{center} 
\caption[]{Summary of discretized Feynman rules using only cubic 
vertices. We have explicitly inserted a factor of $a/m\equiv1/T_0$ for 
each vertex arising from the discretization.} 
\label{NewRules} 
\end{figure} 
 
In this article we shall use this formalism exclusively. In 
Section 2 we illustrate how the cubic vertices 
with fictitious particles reproduce tree diagrams by 
calculating a four gauge boson amplitude, and 
by showing how the $p^+=0$ divergences are resolved 
in tree approximation. 
Next we turn to the 
triangle diagram to one loop order. In Section 3 and 4 we 
calculate the triangle diagram for three off-shell external transverse 
gluons. We extract all of the divergent parts, showing  
how the double logs arising from the entanglement of ultraviolet 
and infra-red divergences in light-cone gauge cancel 
in physical quantities\footnote{In the setup used here,  
the Mandelstam-Leibbrandt (ML) 
trick to avoid this entanglement is {\it not} employed. Indeed, one 
point we wish to stress is that, although the ML prescription 
is convenient for certain purposes, it is by no means 
necessary in a consistent formulation of perturbation 
theory.}  
Concluding remarks are given in Section 5.

\end{section} 
\setcounter{equation}0 
\renewcommand{\theequation}{\thesection.\arabic{equation}} 
\section{Four Gluon Tree Diagrams} 
\label{sec2} 
\begin{figure}[ht] 
\begin{center} 
\begin{tabular}{cccccccc} 
&&&&&&&\\[-.3cm] 
$A_1=$ 
&  
$ 
\begin{array}[c]{c} 
\psfig{file=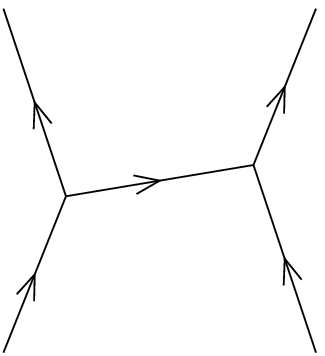,height=1.5cm} 
\end{array} 
$ 
& 
$\quad A_2=$ 
&  
$ 
\begin{array}[c]{c} 
\psfig{file=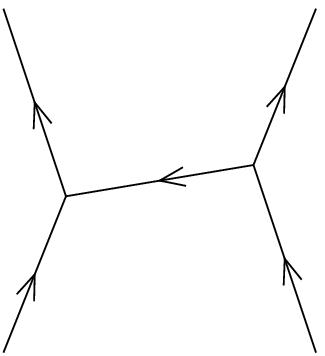,height=1.5cm} 
\end{array} 
$ 
&%\\ 
%&&&\\[-.3cm] 
$\quad A_3=$ 
&  
$ 
\begin{array}[c]{c} 
\psfig{file=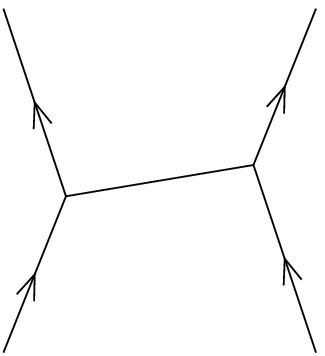,height=1.5cm} 
\end{array} 
$ 
& 
$\quad A_4=$ 
&  
$ 
\begin{array}[c]{c} 
\psfig{file=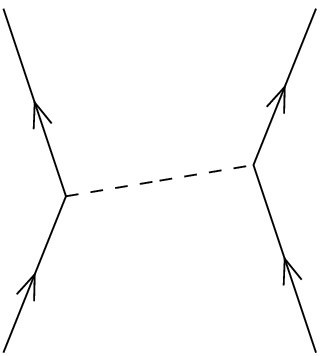,height=1.5cm} 
\end{array} 
$ \\ 
\end{tabular} 
\end{center} 
\caption[]{The four gluon tree diagrams with $t$ 
channel gluon exchange considered in this section.} 
\label{4gluons} 
\end{figure}

Because of the unusual nature of the Feynman rules in our discretized 
light-cone gauge, we 
introduce the reader to this formalism by discussing the tree approximation 
to the two gluon scattering diagrams shown in Fig.~\ref{4gluons}. 
We label the external momenta entering a diagram  
counterclockwise starting from the 
lower left and denote by $k$ the discretized time difference between 
the vertices. Note that $p_i^+=mM_i$. The $t$-channel exchange  
amplitudes for two up gluons to two up gluons associated 
with these diagrams are then given by 
\begin{eqnarray} 
A_1&=&{4g^2\over T_0^2}{-M_1M_3\over 2M_2M_4|M_1-M_4|^3} 
{(M_3p_2^\wedge-M_2p_3^\wedge)(M_1p_4^\vee-M_4p_1^\vee) 
\over e^{(p_1-p_4)^2/2|M_1-M_4|T_0}-1} 
\nonumber\\ 
A_2&=&{4g^2\over T_0^2}{-M_2M_4\over 2M_1M_3|M_1-M_4|^3} 
{(M_3p_2^\vee-M_2p_3^\vee)(M_1p_4^\wedge-M_4p_1^\wedge) 
\over e^{(p_1-p_4)^2/2|M_1-M_4|T_0}-1}\nonumber\\ 
A_3&=&{g^2\over T_0}\sum_{k=1}^\infty f_k  
e^{-k(p_1-p_4)^2/2|M_1-M_4|T_0}{(M_1+M_4)(M_2+M_3) 
\over(M_1-M_4)^2}\nonumber\\ 
A_4&=&{g^2\over T_0}\sum_{k=1}^\infty h_k  
e^{-k(p_1-p_4)^2/2|M_1-M_4|T_0} 
\end{eqnarray} 
Note that we have supplied a wave function factor $e^{akp^+_i}$ 
($e^{-akp^+_i}$)   
for each outgoing (incoming particle), and we  restrict the 
energies by the conservation law $p^+_3+p^+_4=p^+_1+p^+_2$. Also, the 
absolute values account for the different treatment of the case $M_1<M_4$ 
from the case $M_1>M_4$.  
 
These expressions illustrate the 
essential features of the BRT discretization. $A_1$ and $A_2$ show 
the $t$ channel poles due to one gluon exchange: 
\begin{eqnarray} 
{1\over e^{(p_1-p_4)^2/2|M_1-M_4|T_0}-1}\sim {2|M_1-M_4|T_0 
\over(p_1-p_4)^2}. 
\end{eqnarray} 
Notice that time discretization has cut off the large 
$-t=(p_1-p_4)^2$ behavior of the amplitude exponentially. 
Recalling that $T_0=m/a$, we see that sending $a\to0$ 
at fixed $t$ and fixed $m|M_1-M_4|$ just leads to the usual $1/t$ factor of the 
Feynman tree amplitude. Notice also the role of 
$p^+$ discretization in cutting off $p^+=0$ 
singularities. In particular, with both $a,m\neq0$ 
and $t<0$ the amplitudes are strictly zero for $M_1=M_4$, whereas 
after the continuum limit they are infinite at this point.  
With $T_0$ fixed, we can also reach the continuum limit by taking 
all $M_i\to\infty$ as well as $|M_1-M_4|\to\infty$. Either 
way, it is important to keep in mind that the correct 
$p_1^+\to p_4^+$ behavior of the continuum limit is 
only obtained if the continuum limit is first taken with 
$p_1^+\neq p_4^+$. 
 
The expressions for $A_3$ and $A_4$ contain 
the so far undetermined parameters $f_k, h_k$. Since they 
limit the range of the $k$ summation, the formal continuum 
limit just discussed only involves them in the 
combinations $\sum_k f_k$, $\sum_k h_k$, both of 
which are constrained to be unity.  
Thus, in the continuum limit, the sum of all 
four diagrams, $A=A_1+A_2+A_3+A_4$ is given by 
\begin{eqnarray} 
A&=&{g^2\over T_0}\left[ 
-{4p^+_1p^+_3\over p^+_2p^+_4(p^+_1-p^+_4)^2} 
{(p^+_3p_2^\wedge-p^+_2p_3^\wedge)(p^+_1p_4^\vee-p^+_4p_1^\vee)\over (p_1-p_4)^2} 
\right.\nonumber\\  
&& \left. \mbox{} - {4p^+_2p^+_4\over p^+_1p^+_3(p^+_1-p^+_4)^2} 
{(p^+_3p_2^\vee-p^+_2p_3^\vee)(p^+_1p_4^\wedge-p^+_4p_1^\wedge) 
\over (p_1-p_4)^2} + {(p^+_1+p^+_4)(p^+_2+p^+_3)\over(p^+_1-p^+_4)^2}  
+ 1\right]. 
\end{eqnarray} 
It is significant that in the continuum limit no absolute value 
signs are needed.  
 
The individual terms contributing to the continuum limit $A$ show 
quadratic singularities as $p_1^+\to p_4^+$, typical 
of the light-cone gauge. But in the sum, these singularities 
are softened off-shell and disappear on-shell. To show this, 
we note the following identities: 
\begin{eqnarray} 
&&(p^+_3p_2^\wedge-p^+_2p_3^\wedge)(p^+_1p_4^\vee-p^+_4p_1^\vee)=\nonumber\\ 
&&\qquad\qquad 
{1\over2}(p^+_3p_2-p^+_2p_3)\cdot(p^+_1p_4-p^+_4p_1)+{1\over2}(p^+_1-p^+_4)[(p^+_1-p^+_4)(p_2^\wedge p_1^\vee-p_2^\vee p_1^\wedge)\nonumber\\ 
&&\qquad\qquad\qquad\qquad-(p_1-p_4)^\vee(p^+_1p_2^\wedge-p^+_2p_1^\wedge) 
+(p_1-p_4)^\wedge 
(p^+_1p_2^\vee-p^+_2p_1^\vee)]\\ \nonumber\\ 
&&(p^+_3p_2^\vee-p^+_2p_3^\vee)(p^+_1p_4^\wedge-p^+_4p_1^\wedge)=\nonumber\\ 
&&\qquad\qquad 
{1\over2}(p^+_3p_2-p^+_2p_3)\cdot(p^+_1p_4-p^+_4p_1)-{1\over2}(p^+_1-p^+_4) 
[(p^+_1-p^+_4)(p_2^\wedge p_1^\vee-p_2^\vee p_1^\wedge)\nonumber\\ 
&&\qquad\qquad\qquad\qquad-(p_1-p_4)^\vee(p^+_1p_2^\wedge-p^+_2p_1^\wedge) 
+(p_1-p_4)^\wedge(p^+_1p_2^\vee-p^+_2p_1^\vee)]. 
\end{eqnarray} 
Furthermore, by using energy momentum conservation we can rewrite 
the common first term on the r.h.s. of these identities 
\begin{eqnarray} 
&&(p^+_3p_2-p^+_2p_3)\cdot(p^+_1p_4-p^+_4p_1) 
={p^+_1p^+_3+p^+_2p^+_4\over2}(p_1-p_4)^2 +\nonumber\\ 
&&\qquad\qquad\qquad\quad {(p^+_1-p^+_4)^2\over2}(p_1+p_2)^2 
+{p^+_1-p^+_4\over2}[p^+_2p_4^2+p^+_4p_2^2-p^+_1p_3^2-p^+_3p_1^2]. 
\label{invterm} 
\end{eqnarray} 
The contribution of (\ref{invterm}) to the 
continuum limit of $A_1+A_2$ is 
\begin{eqnarray} 
(A_1+A_2)_{\rm I}&=& 
%-{g^2\over T_0}\left[{({p^+_1}^2{p^+_3}^2+{p^+_2}^2{p^+_4}^2) 
%(p^+_1p^+_3+p^+_2p^+_4) 
%\over p^+_1p^+_3p^+_2p^+_4(p^+_1-p^+_4)^2} 
%\right.\nonumber\\&&\qquad\qquad\qquad 
%\mbox{} + {({p^+_1}^2{p^+_3}^2+{p^+_2}^2{p^+_4}^2)(p_1+p_2)^2 
%\over p^+_1p^+_3p^+_2p^+_4(p_1-p_4)^2}\right.\nonumber\\ 
%&&\left. 
%\mbox{} +  
%{({p^+_1}^2{p^+_3}^2+{p^+_2}^2{p^+_4}^2) 
%[p^+_2p_4^2+p^+_4p_2^2-p^+_1p_3^2-p^+_3p_1^2] 
%\over p^+_1p^+_3p^+_2p^+_4(p^+_1-p^+_4)(p_1-p_4)^2}\right]\nonumber\\ 
%&=& 
-{g^2\over T_0}\left[{(p^+_1+p^+_4)(p^+_2+p^+_3)\over(p^+_1-p^+_4)^2}+1 
+{(p^+_1+p^+_2)^2(p^+_1p^+_3+p^+_2p^+_4)\over p^+_1p^+_2p^+_3p^+_4}\right. 
\nonumber\\&&\left. 
\hskip-2.5cm\mbox{} + {({p^+_1}^2{p^+_3}^2+{p^+_2}^2{p^+_4}^2)(p_1+p_2)^2 
\over p^+_1p^+_3p^+_2p^+_4(p_1-p_4)^2} 
+ {({p^+_1}^2{p^+_3}^2+{p^+_2}^2{p^+_4}^2)[p^+_2p_4^2+p^+_4p_2^2-p^+_1p_3^2-p^+_3p_1^2] 
\over p^+_1p^+_3p^+_2p^+_4(p^+_1-p^+_4)(p_1-p_4)^2}\right], 
\label{a1a2I} 
\eea 
and we denote the rest of $A_1+A_2$ by $(A_1+A_2)_{\rm II}$. 
Thus, adding on $A_3+A_4$, we can write the total amplitude 
\bea 
A&=&(A_1+A_2)_{\rm II}-{g^2\over T_0}\left[ 
{(p^+_1+p^+_2)^2(p^+_1p^+_3+p^+_2p^+_4)\over p^+_1p^+_2p^+_3p^+_4} 
+ {({p^+_1}^2{p^+_3}^2+{p^+_2}^2{p^+_4}^2)(p_1+p_2)^2 
\over p^+_1p^+_3p^+_2p^+_4(p_1-p_4)^2}\right. 
\nonumber\\&&\left. 
\hskip1.5cm\mbox{}  
+ {({p^+_1}^2{p^+_3}^2+{p^+_2}^2{p^+_4}^2)[p^+_2p_4^2+p^+_4p_2^2-p^+_1p_3^2-p^+_3p_1^2] 
\over p^+_1p^+_3p^+_2p^+_4(p^+_1-p^+_4)(p_1-p_4)^2}\right]. 
\label{atotal} 
\end{eqnarray} 
We find that $(A_1+A_2)_{\rm II}$ has a continuum limit which  
has no $p^+_1-p^+_4$ denominators: 
\begin{eqnarray} 
  (A_1+A_2)_{\rm II}=-{2g^2\over T_0  }\left[ 
 {(p^+_1 p^+_3+ p^+_2 p^+_4)(p^+_1+p^+_2)p^+_4   
(p_1^\wedge p_2^\vee-p_1^\vee p_2^\wedge) 
\over  p^+_1p^+_3p^+_2p^+_4 (p_1-p_4)^2 }\right. 
\hskip1in&& 
\nonumber\\                          
-\left. {(p^+_1 p^+_3+ p^+_2p^+_4)(p^+_1+p^+_2) 
[(p^+_2p_1^\wedge-p^+_1p_2^\wedge)p_4^\vee 
-(p^+_2p_1^\vee-p^+_1p_2^\vee)p_4^\wedge] 
\over  p^+_1p^+_3p^+_2p^+_4(p_1-p_4)^2}\right]&& 
\end{eqnarray} 
Note that the quadratically singular 
first term in square brackets on the r.h.s. of  
(\ref{a1a2I}) has been exactly 
canceled by the continuum limit of $A_3$, leaving only a linear 
singularity as $p^+_1-p^+_4\to0$, which disappears when all 
of the external gluons are on shell ($p_i^2=0$):  
\begin{eqnarray} 
A_{\rm div}={({p^+_1}^2{p^+_3}^2+{p^+_2}^2{p^+_4}^2)[p^+_2p_4^2+p^+_4p_2^2-p^+_1p_3^2-p^+_3p_1^2] 
\over p^+_1p^+_3p^+_2p^+_4(p^+_1-p^+_4)(p_1-p_4)^2} 
\end{eqnarray} 
This singular contribution has 
opposite signs for $p^+_1>p^+_4$ and $p^+_1<p^+_4$, so we can expect that, 
when this term occurs as a sub-diagram where $M_1-M_4$ is summed, 
a principal value definition of the continuum integral approximation 
to the sum will be appropriate and no divergence will occur. 
It is also interesting to notice 
that when the particles in the initial and final 
state are equally off energy shell, \ie $p_1^2/M_1=p_2^2/M_2$ 
{\it and} $p_3^2/M_3=p_4^2/M_4$, the quantity in square brackets 
becomes $(p^+_1+p^+_2)(p^+_4-p^+_1)(p_3^2/p^+_3+p_1^2/p^+_1)$, with 
no singularity as $p^+_1\to p^+_4$.  
The softening of $p^+=0$ singularities in this 
particular off-shell situation has also been noted in \cite{rozowskyt} 
in the radiative corrections to another four point (branion) process. 
 
Finally, we comment that the on-shell four point 
amplitude assumes a fairly compact form in the Galilei 
center of mass frame ${\bf p}_2=-{\bf p}_1$. In this case 
we obtain 
\begin{eqnarray} 
A^{\rm CM}_{\rm On~shell}&=& 
-{g^2\over T_0}\left[{({p^+_1}^2{p^+_3}^2+{p^+_2}^2{p^+_4}^2)s 
\over p^+_1p^+_3p^+_2p^+_4t} 
%\right.\nonumber\\&&\left.\hskip3cm\mbox{}  
+{(p^+_1+p^+_2)^2(p^+_1p^+_3+p^+_2p^+_4)\over p^+_1p^+_2p^+_3p^+_4} 
\left(                          
1+2{ p_1^\wedge p_4^\vee 
-p_1^\vee p_4^\wedge  
\over  t}\right) 
 \right]\nonumber 
\end{eqnarray} 
where we have introduced the usual Mandelstam invariants 
$s=-(p_1+p_2)^2$ and $t=-(p_4-p_1)^2$. In this form it is 
easy to check that the diagram has the correct value. 
 
For comparison, we record here the value of the diagram 
with $s$ channel poles, also in the Galilei center of mass: 
\bea 
A_s^{\rm CM}&=&-{4g^2\over T_0}{(p_1^++p_2^+)^4p_1^\wedge p_4^\vee 
\over p_1^+p_2^+p_3^+p_4^+ s}\\ 
&=&-{g^2\over T_0}{(p_1^++p_2^+)^4 
\over p_1^+p_2^+p_3^+p_4^+ }\left[{t\over s}+2{p_1^\wedge p_4^\vee- 
p_1^\vee p_4^\wedge\over s}\right]-{g^2\over T_0}{(p_1^++p_2^+)^2(p_1^+p_3^+ 
+p_2^+p_4^+) 
\over p_1^+p_2^+p_3^+p_4^+ } 
\eea 
 
\setcounter{equation}0 
\renewcommand{\theequation}{\thesection.\arabic{equation}} 
 
\begin{section}{Simple Triangle with Three Transverse Gluons} 
 
We will now calculate the one loop three point function 
where all external gluons are transverse. We start with  
a particularly simple process in
which the arrows on all external lines pointing inwards. Since this process 
has no contribution at tree order, it can not 
have ultraviolet divergences. 
In fact, we shall find that the diagram also has no infra-red 
divergences and so is completely finite.\par 
 
\begin{figure}[!ht] 
\begin{center} 
\begin{tabular}{ccc} 
\epsfig{figure=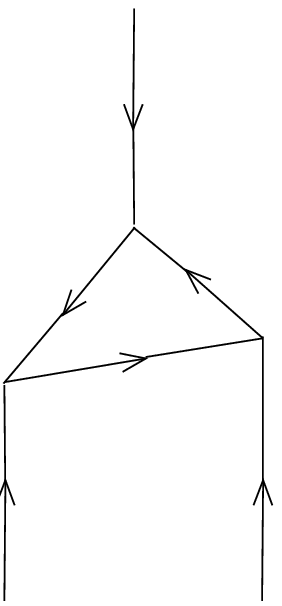,width=1.2cm,height=2cm} & + & 
\epsfig{figure=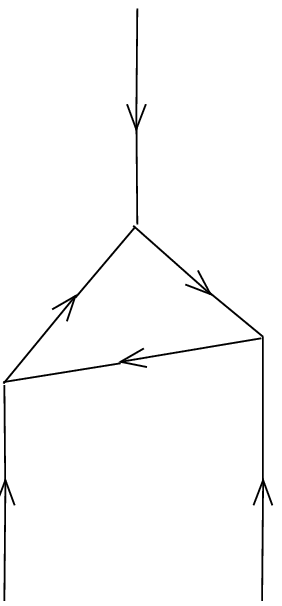,width=1.2cm,height=2cm} 
\end{tabular} 
\caption{Triangle with incoming arrows.} 
\label{botharrowsin} 
\end{center} 
\end{figure} 
 
Using the discretized Feynman rules we see that only the two diagrams 
shown in Fig.~\ref{botharrowsin} contribute to this choice of 
external momenta. One of these is shown in more detail  
in Fig.~\ref{firstarrowsin}. The loop momentum is $q$ and the 
momenta shown are routed along the arrows. This means that the 
momenta $p_1$,$p_2$ and $p_3$ are directed into the vertex and 
momentum conservation reads $\sum p_i=0$. We use the convention that the 
discretized $+$ component of momenta $p_i$ is $m M_i$ and that of $q$ 
is $m l$. Accordingly we have $M_1,M_2>0$ and $M_3=-M_1-M_2<0$. For 
convenience we shall also use $M \equiv -M_3>0$ and $p \equiv 
-p_3>0$ in the following analysis. The discretized times are given by 
$\tau=ak$ and by  
translational invariance we can choose one of the vertices to be at 
$\tau=0$. We then have a transverse loop integral to do, the integral 
over $q^+$ becomes a sum over $l$ and finally we must sum over the 
two discretized times $k_1$ and $k_2$.  
When $k_1>0$ (so also $l>0$), $k_1$ and $k_2^\prime\equiv k_2-k_1$ 
are independently summed from 1 to $+\infty$ and $l$ is summed from 
1 to $M_1-1$. When $k_1<0$ (so also $l<0$), $k_1^\prime\equiv -k_1$ 
and $k_2$ are independently summed from 1 to $+\infty$ and 
$l^\prime\equiv -l$ is summed from 1 to $M_2-1$. 
 
\begin{figure}[!ht] 
\begin{center} 
\psfrag{k0}{\small{$k=0$}} 
\psfrag{k1}{\small{$k=k_1$}} 
\psfrag{k2}{\small{$k=k_2$}} 
\psfrag{p1}{\small{$p_1$}} 
\psfrag{p2}{\small{$p_2$}} 
\psfrag{p3}{\small{$p_3$}} 
\psfrag{q}{\small{$q$}} 
\epsfig{figure=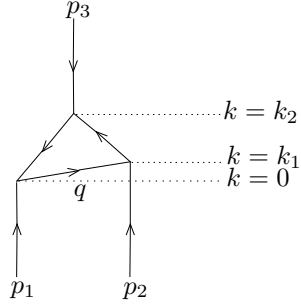,height=4cm} 
\caption{Conventions when calculating the $\Gamma^{\wedge \wedge \wedge}$.} 
\label{firstarrowsin} 
\end{center} 
\end{figure} 
 
By writing out the expression for the two diagrams, shifting the 
transverse loop momentum and simplifying (note that a factor of 
$e^{-akp^-}$ comes from each of the ingoing external lines) we obtain: 
 
\begin{align} 
\Gamma^{\wedge \wedge \wedge} &= \sum_{k,l} \frac{g^3}{T_0^3} 
\frac{2(M_1M_2M_3)^{-1}}{(M_1-l)(M_2+l)|l|} e^{-H}  
\int \frac{d^2 r}{(2 \pi)^3} e^{-\Sigma T \bf{r}^2} 
\left( \frac{T_1}{\Sigma T} K_{23} - M_3r  
\right)^{\wedge}   
\\ 
& \qquad \qquad \qquad \qquad  
\left( \frac{T_2}{\Sigma T} K_{31} - M_1r  
\right)^{\wedge} \left( \frac{T_3}{\Sigma T} K_{12} - M_2r \right)^{\wedge} 
\\ 
&= \left( \frac{g}{T_0} \right)^3 \sum_{k,l} \frac{e^{-H}}{4\pi^2} 
\frac{(M_1M_2M_3)^{-1}}{(M_1-l)(M_2+l)|l|} \frac{T_1T_2T_3}{\Sigma T^4} 
(K_{12}^{\wedge})^3 
\end{align} 
\noindent  
the sum on $k,l$ is constrained as described above and for brevity we 
have made the following definitions:  
 
\begin{eqnarray} 
\label{transktot} 
T_1= \frac{1}{2T_0} \frac{k_1}{l}\;,\qquad 
T_2=\frac{1}{2T_0} \frac{k_2-k_1}{M_2+l}\; ,\qquad  
T_3=\frac{1}{2T_0} \frac{k_2}{M_1-l}. 
\end{eqnarray} 
 
\begin{eqnarray} 
\Sigma T &=& T_1 + T_2 + T_3 \\ 
H &=& \frac{1}{\Sigma T} \left( T_1T_3p_1^2+T_1T_2p_2^2+T_2T_3p_3^2 \right) \\ 
K_{ij} &=& M_i p_j - M_j p_i 
\end{eqnarray} 
 
Note the constraint $lT_1+(M_2+l)T_2-(M_1-l)T_3=0$, which 
implies that for fixed $l$, only two of the $T$'s are 
independent. Also, momentum conservation implies that $K_{ij}$ is 
cyclically symmetric and we therefore use  
$K \equiv K_{12}=K_{23}=K_{31}$.  
 
\par 
 
In general one must take the continuum limit with care because of 
ultraviolet and infra-red divergences that  
might be present. Since we expect no ultraviolet divergences we can 
immediately take $a \to 0$. However, it turns out there are no 
infra-red divergences either and we can see this by  
taking the continuum limit  
$a\to 0$ and $m\to 0$ simultaneously. In the $a \to 0$ limit we can 
replace the sums over $k_1, k_2^\prime$ for $k_1>0$ 
by integrals over $T_1$ and $T_2$ and in the $m \to 0$ limit we can 
replace the sum over $l$ by an integral over $T_3$. The continuous 
transformation between $T_1,T_2,T_3$ and $ak_1,ak_2,ml$ is given by 
(\ref{transktot}) and the jacobian is $\Sigma 
T/4m|l|(mM_1-ml)(mM_2+ml)$. Notice that integrating  
the $T$'s from $0$ to $+\infty$ accounts for summing over the {\em 
whole} range; $l>0,k_1>0$ {\em and} $l<0,k_1<0$. We obtain:

\begin{eqnarray} 
\Gamma^{\wedge \wedge \wedge} &=&  
\left( \frac{g}{T_0} \right) \frac{g^2}{\pi^2} 
\frac{(K^{\wedge})^3}{M_1 M_2 M_3} 
\int dT_1 dT_2 dT_3 
\frac{e^{-H(T_1,T_2,T_3)}T_1T_2T_3}{(T_1+T_2+T_3)^5} 
\\ &=& 
\left( \frac{g}{T_0} \right) \frac{g^2}{\pi^2} 
\frac{(K^{\wedge})^3}{M_1 M_2 M_3} \int_0^\infty dx \int_0^\infty dy 
\frac{xy}{(xp_1^2+xyp_2^2+yp_3^2)(1+x+y)^4} \\ 
\end{eqnarray} 
\noindent 
where we have scaled $x=T_1/T_3$ and $y=T_2/T_3$ and note that 
$H(T_1,T_2,T_3)=T_3 H(x,y,1)$. Notice that for all 
$p_i$'s offshell the double integral converges. For the special 
offshell point $q^2 \equiv p_1^2=p_2^2=p_3^2$ we get: 
 
\begin{equation} 
\Gamma^{\wedge \wedge \wedge} =  
\left( \frac{g}{T_0} \right) \frac{g^2}{\pi^2} 
\frac{(K^{\wedge})^3}{M_1 M_2 M_3} \frac{\chi}{q^2} 
\end{equation}  
\noindent 
where 
 
\begin{equation} 
\chi \equiv \int_0^\infty dx \int_0^\infty dy 
\frac{xy}{(x+xy+y)(1+x+y)^4} \approx 0.030080945 \dots 
\end{equation} 
 
\end{section}

\section{Triangle contributing to charge renormalization} 
\label{sec3} 
We turn now to the main task of this paper, the 
calculation of one-loop diagrams that contribute 
to charge renormalization. The self-energy diagrams 
have been evaluated in \cite{beringrt}, so it remains 
to calculate the vertex corrections, \ie the three 
transverse gluon triangle diagram.   
The kinematics 
of the three gluons are chosen as before.  
We start with the expressions for the diagrams that emerge after 
integrating over the transverse loop momentum. In this section  
we describe the analysis of the remaining 
sums over two discretized times, $k_1a,k_2a$ and one discretized loop  
momentum $p^+=lm$ in the continuum limit, $a,m\to0$. 
 
\begin{figure}[!ht] 
\begin{center} 
\begin{tabular}{|cccc|c|} 
\hline 
\epsfig{figure=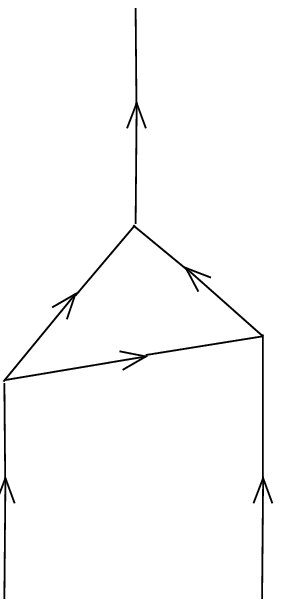,width=1.2cm,height=2cm} & 
\epsfig{figure=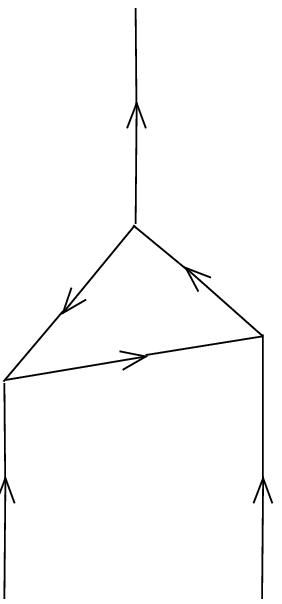,width=1.2cm,height=2cm} & 
\epsfig{figure=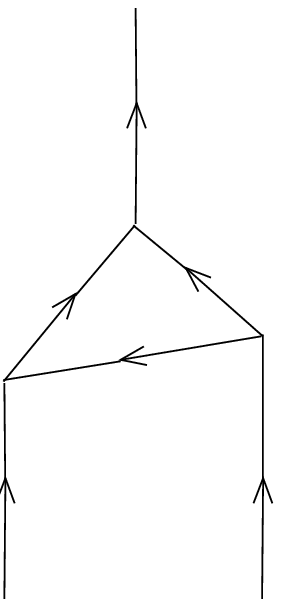,width=1.2cm,height=2cm} & 
\epsfig{figure=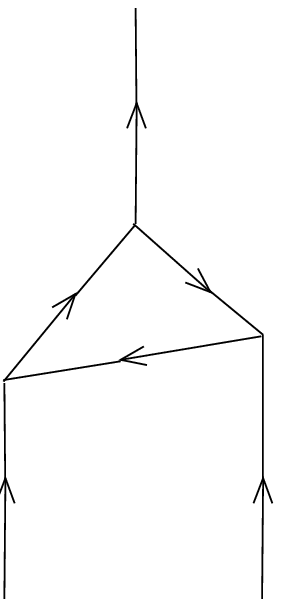,width=1.2cm,height=2cm} & $l>0$ and $k_1>0$ \\ 
\hline 
\epsfig{figure=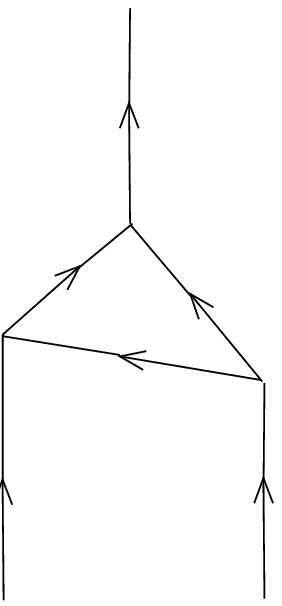,width=1.2cm,height=2cm} & 
\epsfig{figure=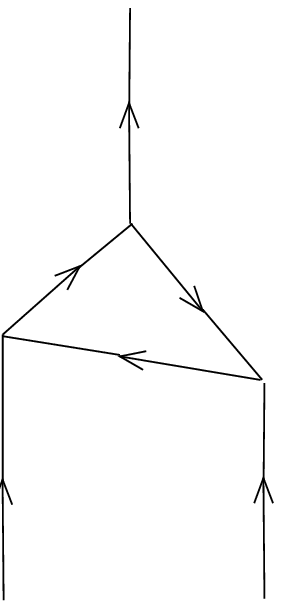,width=1.2cm,height=2cm} & 
\epsfig{figure=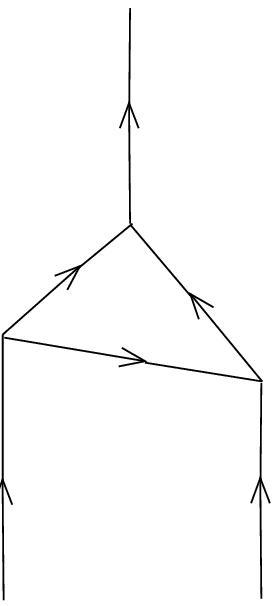,width=1.2cm,height=2cm} & 
\epsfig{figure=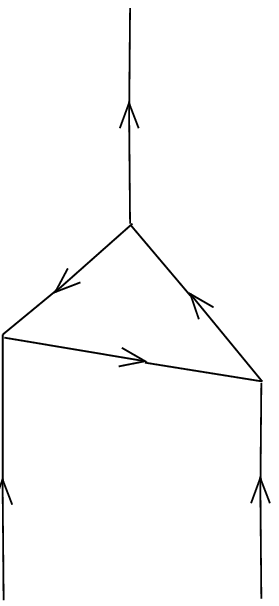,width=1.2cm,height=2cm} & $l<0$ and $k_1<0$ \\ 
\hline 
\end{tabular} 
%\caption{Diagrams with all arrows on internal lines} 
%\label{T6triangles} 
%\end{center} 
%\end{figure*} 
\caption{Diagrams with all internal lines transverse} 
\label{T6triangles} 
\end{center} 
\end{figure} 
\subsection{All Internal Lines Transverse} 
 
We denote the complete three transverse gluon vertex with polarization labels 
$\wedge, \wedge, \vee$ for gluons $1,2,3$  
by $\Gamma^{\wedge\wedge\vee}(p_1,p_2,p_3)$.  
The triangle diagrams displayed in Fig. \ref{T6triangles}, which 
have only transverse gluons on the internal lines,  
produce the following expression 
\begin{eqnarray} 
\Gamma_I^{\wedge\wedge\vee}&=&\frac{g^3K^\wedge}{16\pi^2T_0^3}{M\over M_1M_2} 
\sum_{l,k_1,k_2}\frac{e^{-H}}{|l|(M_2+l)(M_1-l)} 
        \Biggl({\frac{T_1T_2T_3K^2\,A}{M^2(T_1+T_2+T_3)^4} 
   }\nonumber\\ 
&&\hskip2in    +\frac{T_1B_1+T_2B_2+T_3B_3}{(T_1+T_2+T_3)^3} 
\Biggr) 
\end{eqnarray} 
\noindent 
where $K,H$ and the $T_i$'s are as before and 
\begin{eqnarray} 
A&=&\frac{M^2M_1^2}{l^2(M_1-l)^2}+ 
    \frac{M^2M_2^2}{l^2(M_2+l)^2} + 
    \frac{(M_2+l)^2}{(M_1-l)^2}+ 
    \frac{(M_1-l)^2}{(M_2+l)^2 }\\ 
B_1&=&\frac{M_2M_1^3}{l^2(M_1-l)^2}+\frac{M_1M_2^3}{l^2(M_2+l)^2} \\ 
B_2&=&-\frac{MM_2^3}{l^2(M_2+l)^2}-\frac{M_2(M_2+l)^2}{M(M_1-l)^2} 
    -\frac{M_2(M_1-l)^2}{M(M_2+l)^2}\\ 
B_3&=&-\frac{MM_1^3}{l^2(M_1-l)^2}-\frac{M_1(M_2+l)^2}{M(M_1-l)^2} 
    -\frac{M_1(M_1-l)^2}{M(M_2+l)^2} 
\end{eqnarray} 
The triangles with only two transverse gluon 
internal lines, 
denoted $\Gamma^{\wedge\wedge\vee}_{II}$ will be dealt with in the following subsection. 
We introduce the following 
notation that will help streamline some of the formulae: $P_i^*\equiv 
p_i^2/M_i,\ P^*\equiv p^2/M=-P_3^*$. For example,  
\begin{eqnarray} 
K^2=-M_1M_2M_3(P_1^*+P_2^*+P_3^*)=M_1M_2M(P_1^*+P_2^*-P^*). 
\end{eqnarray}  
 
The vertex function should be antisymmetric under the interchange 
$p_1, M_1\leftrightarrow p_2, M_2$. In view of the explicit 
overall factor of $K^\wedge$ which is odd under this transformation, 
the coefficient of $K^\wedge$ should be even. Inspection of 
the above expression for the triangle diagram shows that this 
symmetry is realized in the following way. The expression for the 
summation range $k_1=-k^\prime_1, l=-l^\prime<0$ is precisely 
equal to that for the range $k_1, l >0$ {\it with the 
interchange of variables $p_1, M_1\leftrightarrow p_2, M_2$}. 
Thus, it is only necessary to explicitly calculate for 
one time ordering, say $k_1, l > 0$. Then adding to this the 
result of the interchange gives the complete answer. 
 
We are now dealing with potentially ultraviolet divergent diagrams.  
To reveal the ultraviolet structure we 
consider the continuum limit in the order $a\to 0$ followed by 
$m\to 0$. Recall that $a\neq0$ serves as our ultraviolet cutoff. 
In the $a\to0$ limit we can attempt to replace the 
sums over $k_1, k_2^\prime$ ($k_1^\prime, k_2$) for $k_1>0$ ($k_1<0$) 
by integrals over $T_1$ and $T_2$ ($T_3$) just as in the previous 
section. Since we wish to keep $l$  
fixed in this first step, for the case 
$k_1>0$ we express $T_3$ 
in terms of $T_1$ and $T_2$: $T_3=(lT_1+(M_2+l)T_2)/(M_1-l)$. 
For the case $k_1<0$, it is more convenient to express $T_2$ 
in terms of $T_1$ and $T_3$: $T_2=(l^\prime T_1 
+(M_1+l^\prime)T_3)/(M_2-l^\prime)$. We  
find $\Sigma T=(MT_2+M_1T_1)/(M_1-l)=(MT_3+M_2T_1)/(M_2-l^\prime)$. 
For the $A$ term, 
this procedure encounters no obstacle, and we obtain 
(displaying explicitly the contribution for $k_1,l>0$) 
\begin{eqnarray} 
\Gamma_A^{\wedge\wedge\vee}&\to&\frac{g^3K^\wedge}{4\pi^2T_0}{M\over M_1M_2} 
\left\{\sum_{l}\int dT_1dT_2 
         {\frac{(M_1-l)^2T_1T_2(lT_1+(M_2+l)T_2)K^2\,A}{M^2(M_1T_1+MT_2)^4}}  
e^{-H(T_1,T_2)} \; \right.\nonumber\\ 
&&\hskip2in \left.+ \; (\;1\leftrightarrow\; 2\;)\phantom{\int}\right\}\nonumber\\ 
&=&\frac{g^3K^\wedge}{4\pi^2T_0}{M\over M_1M_2} 
\left\{\sum_{l}\int dT 
         {\frac{(M_1-l)^2T(lT+(M_2+l))K^2\,A}{M^2H(T,1)(M_1T+M)^4}} 
 \; + \; (\;1\leftrightarrow\; 2\;)\right\}\; .  
\end{eqnarray} 
It will be useful to note that $H$ can be written in the  
alternative forms 
\begin{eqnarray} 
H 
%&\equiv&\frac{T_1T_3p_1^2+T_1T_2p_2^2+T_2T_3p_3^2}{T_1+T_2+T_3} 
%\nonumber\\ 
&=&(M_2+l)T_2P^* 
+lT_1P_1^*+{K^2\over M_1M}\left[ 
{(M_1-l)T_1T_2\over MT_2+M_1T_1}\right]\\ 
&=&(M_1+l^\prime)T_3P^* 
+l^\prime T_1P_2^*+{K^2\over M_2M}\left[ 
{(M_2-l^\prime)T_1T_3\over MT_3+M_2T_1}\right], 
\end{eqnarray} 
where the first is useful when $k_1,l>0$ and the second for 
$k_1,l<0$.  
\iffalse 
&=&(M_2+l)T_2\left({p_1^2\over M_1}+{p_2^2\over M_2}\right)+lT_1{p_1^2\over M_1}-{K^2T_2\over M_1M_2}\left[ 
{lT_1+(M_2+l)T_2\over MT_2+M_1T_1}\right]\\ 
&=&(M_1+l^\prime)T_3\left({p_1^2\over M_1} 
+{p_2^2\over M_2}\right)+l^\prime T_1{p_2^2\over M_2} 
-{K^2T_3\over M_1M_2}\left[ 
{l^\prime T_1+(M_1+l^\prime)T_3\over MT_3+M_2T_1}\right] \\ 
\fi 
 
However the $B$ terms produce logarithmically divergent 
integrals with this procedure, so they must be handled 
differently. 
To deal with these logarithmically divergent terms, we first note 
the identities: 
\begin{eqnarray} 
{T_1\over(T_1+T_2+T_3)^3}&=&-{\partial\over\partial T_2} 
{(M_1-l)^3\over2M}{T_1\over(MT_2+M_1T_1)^2}\\ 
&=&-{\partial\over\partial T_3} 
{(M_2-l^\prime)^3\over2M}{T_1\over(MT_3+M_2T_1)^2}\\ 
{T_2\over(T_1+T_2+T_3)^3}&=&-{\partial\over\partial T_2} 
{(M_1-l)^3\over2M^2}{M_1T_1+2MT_2\over(MT_2+M_1T_1)^2}\\ 
{T_3\over(T_1+T_2+T_3)^3}&=&-{\partial\over\partial T_3} 
{(M_2-l^\prime)^3\over2M^2}{M_2T_1+2MT_3\over(MT_3+M_2T_1)^2}, 
\end{eqnarray} 
where the partial derivatives are taken with $T_1$ fixed. 
 
Because of the divergences we can't immediately write the 
continuum limit of the $B$ terms as an integral. However 
we can make the substitution $e^{-H}\to (e^{-H}-e^{-H_0})+e^{-H_0}$, 
where $H_0$ is chosen to be an appropriate simplified 
version of $H$, which coincides with $H$ at $T_2=0$.  
For $k_1,l>0$, it is convenient to choose 
$H_0=(lT_1+(M_2+l)T_2)P_1^*$, whereas for $k_1,l<0$ 
$H^\prime_0=(l^\prime T_1+(M_1+l^\prime)T_3)P_2^*$ is more convenient. 
Then the factor $(e^{-H}-e^{-H_0})$ 
regulates the integrand at small $T_i$ so that the sums may then 
safely be replaced by integrals. We shall denote the contributions from these 
terms by $\Gamma^{\wedge\wedge\vee}_{B1}$. 
Then using the above identities, an 
integration by parts (for which the surface term 
vanishes) makes the integrand similar to that in $\Gamma^{\wedge\wedge\vee}_A$ 
and simplifications can be achieved. For details see Appendix A.1. 
\begin{eqnarray} 
\Gamma^{\wedge\wedge\vee}_A 
+\Gamma^{\wedge\wedge\vee}_{B1}&\to&{g^3K^\wedge\over4\pi^2 T_0}{M\over M_1M_2} 
\left\{\sum_{l=1}^{M_1-1}\int_0^\infty dT% 
\left[{TK^2(M_1-l)^2A^\prime\over M^2H(M+M_1T)^3}\right.\right.\nonumber\\ 
&&\left.\left.+{lT(H_0-H)(M_1-l)M_1A^\prime 
\over M^2H(lT+M_2+l)(M+M_1T)} 
%\right.\right.\nonumber\\&&\hskip1.5in\left.\left.\mbox{} 
-{lT(H_0-H)(M_1-l)M_1A\over M^2H(M+M_1T)^2}\right] 
\right.\nonumber\\&&\hskip3in\left.\mbox{}\phantom{\int_)^\infty} 
\; + \; (\;1\leftrightarrow\; 2\;)\right\} , 
\label{ab1} 
\end{eqnarray} 
\noindent 
where: 
\begin{eqnarray} 
A^\prime &=& {M^2M_2^2\over lM_1(M_2+l)^2}-{M_2(M_2+l)^2\over M_1M(M_1-l)} 
-{M_2(M_1-l)^3\over M_1M(M_2+l)^2}\ . 
\end{eqnarray} 
Since the integrand of (\ref{ab1}) is a  
rational function of $T$ the last integral 
can also be done. We sketch the evaluation in Appendix B. 
 
There remains the contribution of the term $e^{-H_0}$ which would 
give a divergent integral. However, because  the $T_1,T_2$ ($T_1, T_3$) 
dependence in the exponential is disentangled by our choice of $H_0$, the  
sums can be directly analyzed in the $a\to0$ limit, giving 
an explicit expression for the divergent part in terms of the lattice 
cutoff. We denote this contribution, containing the 
ultraviolet divergence of the triangles, by $\Gamma^{\wedge\wedge\vee}_{B2}$. Referring to 
Appendix A.2 for details we obtain: 
\begin{eqnarray} 
\Gamma_{B2}^{\wedge\wedge\vee}&=&\frac{g^3K^\wedge}{8\pi^2T_0}{M\over M_1M_2}  
\left\{\left[\sum_{l=1}^{M_1-1}{M_1-l\over MM_1}\left({N_1l\over M_1} 
+{N_2(M_2+l)\over M}\right)\left(\ln{2p_1^+\over ap_1^2} 
+f\left({\alpha\over\beta}\right)\right)\right.\right.\nonumber\\ 
&&\left.\left.-\sum_{l=1}^{M_1-1}{M_1-l\over MM_1}\left({N_1l\over M_1} 
-{N_2(M_2+l)\over M}\right) 
{\alpha\over\beta}f^\prime\left({\alpha\over\beta}\right)\right] 
+ \quad (\;1\; \leftrightarrow\; 2\;)\quad\right\} 
\nonumber\\ 
&=&-\frac{g^3K^\wedge}{4\pi^2T_0}{M\over M_1M_2}  
\left\{\sum_{l=1}^{M_1-1}\left[{B^\prime\over M} 
\left(\ln{2p_1^+\over ap_1^2} 
+f\left({\alpha\over\beta}\right)\right)+{AM_2(M_1-l)^2\over M^3M_1} 
{\alpha\over\beta}f^\prime\left({\alpha\over\beta}\right)\right]\right. 
\nonumber\\ 
&&\hskip3in\left.+ \quad (\;1\; \leftrightarrow\; 2\;) 
\phantom{{\alpha\over\beta}}\right\}, 
\end{eqnarray} 
where we have defined 
\begin{eqnarray} 
B^\prime= 
{(M_1-l)^3\over M^2(M_2+l)}+{MM_1\over l(M_1-l)} 
+{(M_2+l)^3\over M^2(M_1-l)}\; . 
\end{eqnarray} 
and 
\begin{eqnarray} 
N_1 &\equiv& \frac{B_1(M_1-l)}{l}+B_3, \quad N_2 \equiv 
\frac{B_2(M_1-l)}{M_2+l}+B_3 \\ 
f(x) &=& \frac{\ln x}{1-x}-x \int_0^\infty dt e^{-xt} 
\frac{1-xt-e^{-xt}}{(1-e^{-xt})^2} \ln (1-e^{-t}) \\ 
\alpha &\equiv& \frac{M_1}{l}, \quad \quad \beta \equiv \frac{M}{M_2+l} 
\end{eqnarray} 
 
In $\Gamma^{\wedge\wedge\vee}_{B2}$ we can further simplify the term proportional to  
$\ln(2p^+_1/ap_1^2)$, which contains the ultraviolet divergence 
of the triangle diagrams. We obtain  
\begin{eqnarray} 
\Gamma_{\rm div}^{\wedge\wedge\vee}&=&-\frac{g^3K^\wedge}{4\pi^2T_0}{M\over M_1M_2}  
\left\{\sum_{l=1}^{M_1-1}{1\over M}\left[ 
{(M_1-l)^3\over M^2(M_2+l)}+{MM_1\over l(M_1-l)} 
+{(M_2+l)^3\over M^2(M_1-l)} 
\right]\ln{2p_1^+\over ap_1^2}\right.\nonumber\\ 
&& \hskip2in\left.\phantom{{M_1\over3}}  
+ \quad (\;1\; \leftrightarrow\; 2\;)\quad\right\} 
\nonumber\\ 
&=& -\frac{g^3K^\wedge}{4\pi^2T_0}{M\over M_1M_2}\left\{\ln{2p_1^+\over ap_1^2} 
\left[\psi(M_1+M_2)-\psi(M_2+1)+3\psi(M_1)+3\gamma 
\phantom{{M_1\over3}}\right.\right. 
\nonumber\\ 
&&\left.\left. + {M_1-1\over M^3} 
\left(-{11\over3}M_1^2-7M_1M_2-4M_2^2+{M_1\over3}\right)\right] 
+\quad (\;1\; \leftrightarrow\; 2\;)\quad\right\}. 
\end{eqnarray} 
\noindent 
where $\psi$ is the digamma function. 
 
%\begin{equation} 
%\psi (x) = \frac{d}{dx} \ln (\Gamma(x)). 
%\end{equation} 
 
Writing out the 
terms from interchanging $1\leftrightarrow 2$ in this expression, 
and simplifying we obtain 
%%%% 
%\newpage 
%%%% 
\begin{eqnarray} 
\Gamma^{\wedge\wedge\vee}_{\rm uv}&=&\frac{g^3K^\wedge}{16\pi^2T_0} 
{M\over M_1M_2}\left\{\left[\ln{2p_1^+\over ap_1^2} +\ln{2p_2^+\over ap_2^2} 
\right] 
\left[-4\left(\psi(M)+\psi(M_1)+ 
\psi(M_2)+3\gamma\right)\phantom{{8\over M}} 
\right.\right.\nonumber\\ 
&&\left. 
\hskip2in+2\left({11\over3}-{8\over M}+{M\over M_1M_2} 
+{2M_1M_2\over M^3}+{1\over3M^2}\right)\right]\nonumber\\ 
&& \left.\mbox{} - \ln{p_1^+p_2^2\over p_+p_1^2}\left[ 
8(\psi(M_1)-\psi(M_2))+{2(M_1-M_2)\over 3M^3} 
(-11M^2+2M_1M_2-1)-{M_1-M_2\over M_1M_2}\right]\right\}\nonumber\\ 
&\to&\frac{g^3K^\wedge}{16\pi^2T_0}{M\over M_1M_2} 
\left\{2\sum_{i=1}^3\ln{2|p_i^+|\over ap_i^2}\left[-4\left(\ln|M_i| 
+\gamma\right)+{22\over9}\right]\right.\nonumber\\&& 
+\left.\left[\ln{p^2p_1^+\over p^+p_1^2} +\ln{p^2p_2^+\over p^+p_2^2} 
\right] 
\left[-4\left(\ln M 
+\gamma\right)+{22\over9}\right] 
-4\left(\ln {M_1\over M_2}\right)\ln{p_1^2p_2^+\over p_1^+p_2^2} 
\right.\nonumber\\ 
&& \left.\mbox{} - \ln{p_1^+p_2^2\over p_2^+p_1^2}\left[ 
8\ln{p^+_1\over p^+_2}+{2(p^+_1-p^+_2)\over 3p^+} 
\left(-11+{2p^+_1p^+_2\over p^{+2}}\right)\right]\right\} . 
\end{eqnarray} 
where the final expression, valid at large $M, M_1, M_2$, has 
been arranged so that the $uv$ divergence appears symmetrically 
among the three legs of the vertex. 
 
Comparing to the zeroth order vertex, $-2gK^\wedge M/M_1M_2T_0$, 
we see that the ultraviolet divergence of the triangle 
is contained in the multiplicative factor 
\begin{eqnarray} 
1+\frac{g^2}{16\pi^2}\ln{2\over a}\left[4\left(\ln M+\ln M_1+ 
\ln M_2+3\gamma\right)-{22\over3}\right]. 
\label{triangleren} 
\end{eqnarray} 
To see that this is actually the correct renormalization factor, recall that  
the self energy calculation of \cite{beringrt} implies the 
gluon wave function renormalization factor 
\begin{eqnarray}Z(Q)=1-\frac{g^2N_c}{16\pi^2}\left\{ 
\left[8(\ln M +\gamma)-{22\over3}\right]\ln{2Q^+\over aQ^2}+{4\over3}\right\}. 
\end{eqnarray} 
Thus the appropriate wave function renormalization 
factor for the triangle,  $\sqrt{Z(p_1)Z(p_2)Z(p)}$, contains the 
ultraviolet divergent factor  
\begin{eqnarray}1-\frac{g^2N_c}{16\pi^2} 
\left[4(\ln MM_1M_2 +3\gamma)-{11}\right]\ln{2\over a}, 
\end{eqnarray} 
so the divergence for the renormalized triangle is contained 
in the multiplicative factor  
\begin{eqnarray} 
1+{11\over3}\frac{g^2N_c}{16\pi^2}\ln{2\over a} 
%\right.\nonumber\\  
%&&\left.-\left(4\ln M-{11\over3}\right)\ln{p^+\over p^2} 
%-\left(4\ln M_1-{11\over3}\right)\ln{p_1^+\over p_1^2} 
%-\left(4\ln M_2-{11\over3}\right)\ln{p_2^+\over p_2^2}-2\right], 
\end{eqnarray} 
implying the correct relation of renormalized to bare charge 
\begin{eqnarray} 
\label{charge} 
g_R=g\left(1+{11\over24\pi}\alpha_sN_c\ln{2\over a}\right), 
\end{eqnarray} 
where $\alpha_s=g^2/2\pi$. 
 
Putting everything together,  
the amplitude for the triangle with only transverse internal 
lines is given in the continuum limit by: 
\begin{eqnarray} 
&&\hskip-.2in\Gamma_I^{\wedge\wedge\vee}=\frac{g^3K^\wedge}{4\pi^2T_0} 
{M\over M_1M_2} 
\left\{{1\over M_1}\sum_{l=1}^{M_1-1} 
\left[\int_0^\infty dT I_1  
+S_1\right] 
 \mbox{} +\quad(\; 1\; \leftrightarrow \; 2\;) 
\right\}\nonumber\\ 
&&\hskip1in+\frac{g^3K^\wedge}{16\pi^2T_0}{M\over M_1M_2} 
\left\{2\sum_{i=1}^3\ln{2|p_i^+|\over ap_i^2}\left[-4\left(\ln|M_i| 
+\gamma\right)+{22\over9}\right]\right.\nonumber\\&& 
\hskip1in+\left.\left[\ln{p^2p_1^+\over p^+p_1^2} +\ln{p^2p_2^+\over p^+p_2^2} 
\right] 
\left[-4\left(\ln M 
+\gamma\right)+{22\over9}\right] 
\right.\nonumber\\ 
&&\hskip1in \left.\mbox{} - \ln{p_1^+p_2^2\over p_2^+p_1^2}\left[ 
4\ln{p^+_1\over p^+_2}+{2(p^+_1-p^+_2)\over 3p^+} 
\left(-11+{2p^+_1p^+_2\over p^{+2}}\right)\right]\right\} 
\label{finaltriangle} 
\end{eqnarray} 
where 
\begin{eqnarray} 
I_1&=&I\left(T,p_1,p_2,{l\over M_1}\right) \equiv  
\left.{\frac{M_1(M_1-l)^2TK^2\,A^\prime}{M^2H(T,1)(M_1T+M)^3}} 
+{lT(H_0-H)(M_1-l)M_1^2A^\prime 
\over M^2H(lT+M_2+l)(M+M_1T)} 
\right.\nonumber\\&&\hskip2in\left. 
-{lT(H_0-H)(M_1-l)M_1^2A\over M^2H(M+M_1T)^2}\right.\\ 
S_1&=&S\left(M_1,M_2, {l\over M_1}\right)\equiv -\left[{M_1B^\prime\over M} 
f\left({\alpha\over\beta}\right) + {AM_2(M_1-l)^2\over M^3} 
{\alpha\over\beta}f^\prime\left({\alpha\over\beta}\right)\right]\\ 
&&I_2=I\left(T,p_2,p_1,{l\over M_2}\right),\qquad  
S_2=S\left(M_2,M_1, {l\over M_2}\right) 
\end{eqnarray} 
\iffalse 
+\sum_{l=1}^{M_1-1}\frac{M_1-l}{|l|(M_2+l)} 
\sum_{k_1,k^\prime_2}  
\left[{\frac{k_1N_1+k^\prime_2N_2}{(k_1\alpha 
+k^\prime_2\beta)^3}}\right]u_1^{k_1+k_2^\prime} 
-\left[{B^\prime\over M_1T+M}  
+ {(M_1-l)C^\prime\over(M_1T+M)^2}\right]{T(H_0-H)\over (lT+M_2+l)H} 
\fi 
and where we recall, for convenience, our definitions (appropriate to  
the case $k_1,l>0$) 
\begin{eqnarray} 
A&=&\frac{M^2M_1^2}{l^2(M_1-l)^2}+ 
    \frac{M^2M_2^2}{l^2(M_2+l)^2} + 
    \frac{(M_2+l)^2}{(M_1-l)^2}+ 
    \frac{(M_1-l)^2}{(M_2+l)^2 }\nonumber\\ 
A^\prime &=& {M^2M_2^2\over lM_1(M_2+l)^2} 
-{M_2(M_2+l)^2\over M_1M(M_1-l)} 
-{M_2(M_1-l)^3\over M_1M(M_2+l)^2}\nonumber\\ 
B^\prime &=&{(M_2+l)^3\over M^2(M_1-l)}+{(M_1-l)^3\over 
M^2(M_2+l)}+{MM_1\over l(M_1-l)}\nonumber\\ 
H&=&H(T,1)=(M_2+l)P_3^* 
+lTP_1^*+ 
{(M_1-l)T\over M+M_1T}{K^2\over M_1M}\nonumber\\ 
H_0&=&H_0(T,1)=(M_2+l)P_1^* 
+lTP_1^*\nonumber\\ 
{\alpha\over\beta}&=& {M_1(M_2+l)\over lM}\; . 
\end{eqnarray} 
 
To complete the continuum limit we assume $M, M_1, M_2$ large 
and attempt to replace 
the sums over $l$ by integrals over a continuous variable 
$\xi=l/M_1$, with $0<\xi<1$. This procedure is obstructed 
by singular behavior of the integrand for $\xi$ near 0 or 1. 
When this occurs, we introduce a cutoff $\epsilon<<1$, 
and only do the replacement for $\epsilon<\xi<1-\epsilon$, 
dealing with the sums directly in the singular regions. 
The detailed analysis is presented in Appendix A. 
Referring to Eq.~\ref{12intsum1sum2}, we see that we can write 
\begin{eqnarray} 
&&\hskip-.2in\Gamma_I^{\wedge\wedge\vee}= 
\Gamma_{I,{\rm finite}}^{\wedge\wedge\vee} 
+\frac{g^3K^\wedge}{16\pi^2T_0}{M\over M_1M_2} 
\left\{2\sum_{i=1}^3\ln{2|p_i^+|\over ap_i^2}\left[-4\left(\ln|M_i| 
+\gamma\right)+{22\over9}\right]\right.\nonumber\\&& 
\hskip1in+\left.\ln{p^2p_1^+\over p^+p_1^2} 
\left[-4\ln {M\over M_1}+{22\over9}\right] +\ln{p^2p_2^+\over p^+p_2^2} 
\left[-4\ln {M\over M_2}+{22\over9}\right]\right\}\nonumber\\ 
&&+\frac{g^3K^\wedge}{4\pi^2T_0}\left[ 
{2\pi^2\over3}+ {M\over M_1M_2}\left[(\ln M_1 
+\gamma)\left\{\left({M_1p^2+Mp_1^2\over M_1p^2-Mp_1^2} 
+{M_1p_2^2-M_2p_1^2\over M_1p_2^2+M_2p_1^2 }\right) 
\ln{M_1p^2\over Mp_1^2}-3+{\pi^2\over6} \right\} \right.\right.\nonumber\\ 
&&\left.\left. +(\ln M_2 
+\gamma)\left\{\left({M_2p^2+Mp_2^2\over M_2p^2-Mp_2^2} 
-{M_1p_2^2-M_2p_1^2\over M_1p_2^2+M_2p_1^2 }\right) 
\ln{M_2p^2\over Mp_2^2}-3+{\pi^2\over6} \right\}  \right]\right] 
\label{finaltriangle1} 
\end{eqnarray} 
 
\subsection{Longitudinal Internal Gluons} 
\begin{figure*}[!ht] 
\begin{center} 
\begin{tabular}{|cccccc|c|} 
\hline 
\epsfig{figure=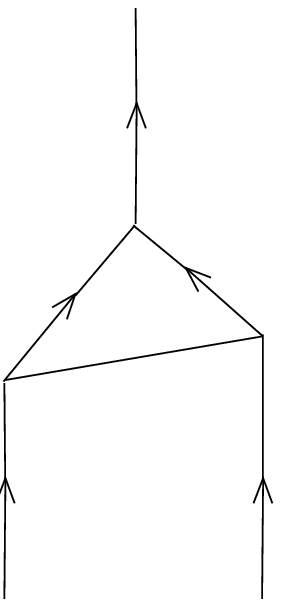,height=2cm,width=1.2cm} & 
\epsfig{figure=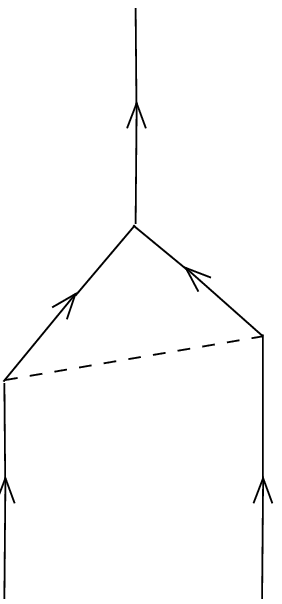,height=2cm,width=1.2cm} & 
\epsfig{figure=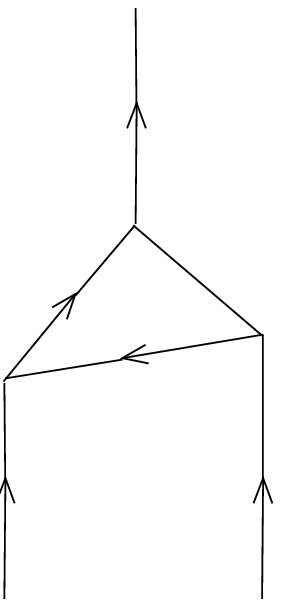,height=2cm,width=1.2cm} & 
\epsfig{figure=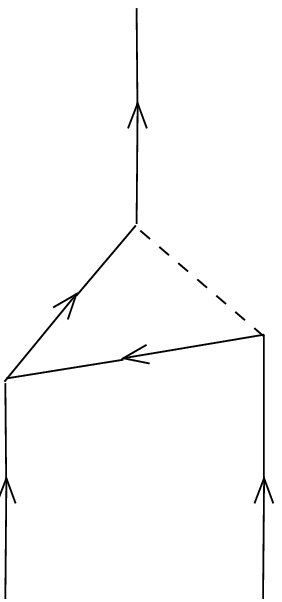,height=2cm,width=1.2cm} & 
\epsfig{figure=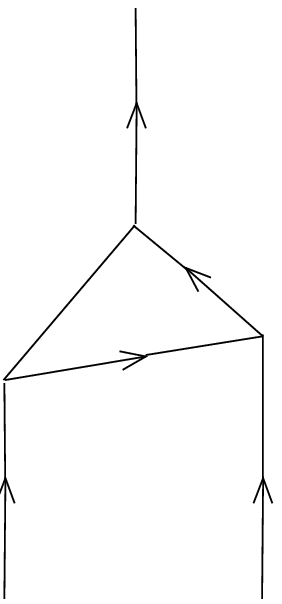,height=2cm,width=1.2cm} & 
\epsfig{figure=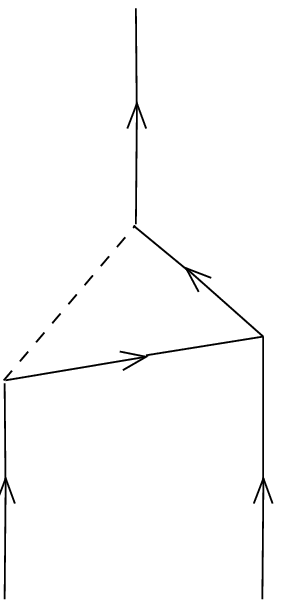,height=2cm,width=1.2cm} & $l>0$ and $k_1>0$ \\ 
\hline 
\epsfig{figure=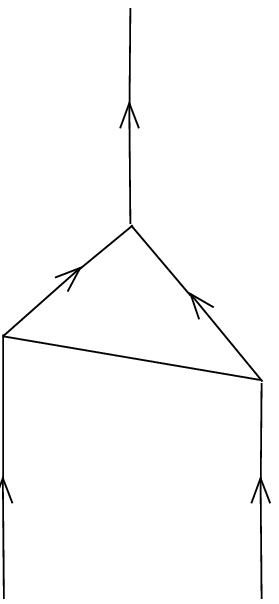,height=2cm,width=1.2cm} & 
\epsfig{figure=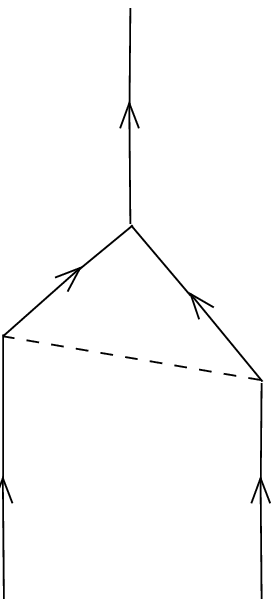,height=2cm,width=1.2cm} & 
\epsfig{figure=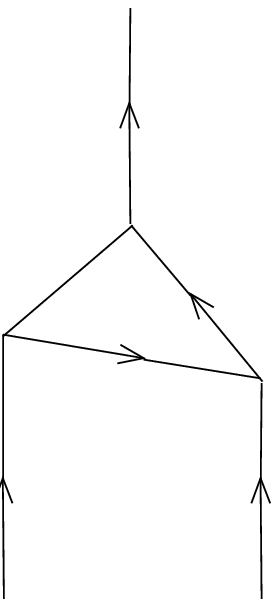,height=2cm,width=1.2cm} & 
\epsfig{figure=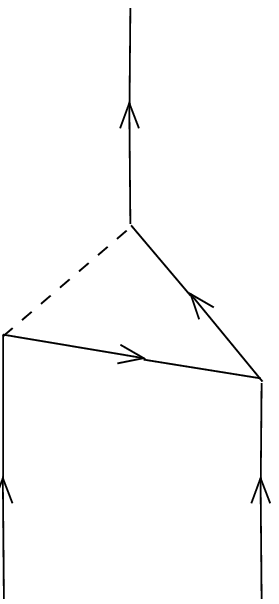,height=2cm,width=1.2cm} & 
\epsfig{figure=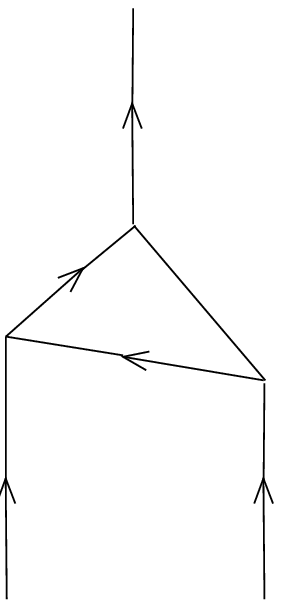,height=2cm,width=1.2cm} & 
\epsfig{figure=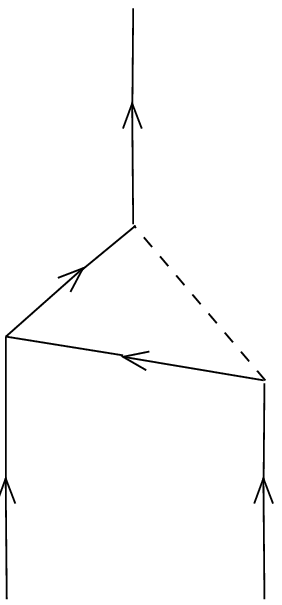,height=2cm,width=1.2cm} & $l<0$ and $k_1<0$ \\ 
\hline 
\end{tabular} 
\caption{Diagrams with one solid or dashed internal line} 
\label{straightdashed} 
\end{center} 
\end{figure*}

In addition to the diagrams with three transverse gluon internal 
lines discussed in 
the previous subsection, there are diagrams where one of the 
internal lines is a longitudinal gluon (solid line with no 
arrow) or a fictitious gluon, whose exchange represents the 
four gluon vertex as the concatenation of two cubic vertices 
(dashed line). The various possibilities are shown in  
Fig.~\ref{straightdashed}.  
 
Again we start with the expression for the sum of these diagrams obtained after 
doing the transverse momentum integrals: 
\begin{eqnarray} 
\Gamma^{\wedge\wedge\vee}_{II}&=&-\frac{g^3}{16\pi^2T_0^2}K^\wedge 
\sum_l \frac{1}{l(M_2+l)(M_1-l)} 
        e^{-H} \nonumber\\ 
&&    \Biggl( \frac{T_1}{(T_1+T_2+T_3)^2}\,C_1+ 
    \frac{T_2}{(T_1+T_2+T_3)^2}\,C_2+ 
    \frac{T_3}{(T_1+T_2+T_3)^2}\,C_3 
    \Biggr)\; , 
\end{eqnarray} 
where: 
\begin{eqnarray} 
C_1 &=& f_{k_1}\frac{M_3(2M_1-l)(2M_2+l)}{l(M_1-l)(M_2+l)} +  
	h_{k_1}\frac{M_3l}{(M_2+l)(M_1-l)} \\ 
C_2 &=& f_{k_2-k_1}\frac{(M_2-l)(M_3+M_1-l)(M_1-l)}{|l|(M_2+l)M_1} + 
	h_{k_2-k_1}\frac{(M_1-l)(M_2+l)}{M_1|l|} \\ 
C_3 &=& -f_{k_2}\frac{(M_1+l)(M_3+M_2+l)(M_2+l)}{|l|(M_1-l)M_2} - 
	h_{k_2}\frac{(M_2+l)(M_1-l)}{|l|M_2} 
\end{eqnarray} 
Since the $f_k, h_k$ are assumed to fall off rapidly with $k$, 
the corresponding sum can never be replaced by an integral.  
Consider first the term with $C_1$. Then for $k_1>0(<0)$ 
only the $k_2^\prime$ ($k_2$) sum ranges freely from 1 
to $\infty$. writing out $H(T_1,T_2)$ explicitly in terms of 
$k_1,k_2^\prime$, 
\begin{eqnarray} 
H={p^2k_2^\prime\over 2MT_0}+{p_1^2k_1\over 2M_1T_0} 
+{K^2k_1\over2T_0MM_1}{(M_1-l)k_2^\prime\over Mlk_2^\prime+M_1k_1(M_2+l)}, 
\end{eqnarray} 
we see that, due to the limited range of $k_1$, the only term that 
can get large is the first one. Furthermore all the 
other terms stay of order $O(a)$ in the limit $a\to0$ at 
fixed $m$, which we are studying. Thus, writing $u=e^{-p^2/2MT_0}$, 
we see that we require the sum 
\begin{eqnarray} 
\sum_{k^\prime_2=1}^\infty{u^{k^\prime_2} 
\over(k^\prime_2+z)^2}= \psi^\prime(z+1) +O((1-u)\ln(1-u)) 
\label{c1sum} 
\end{eqnarray} 
in the $a\to0$ limit. Thus the $C_1$ contribution has the 
$a\to0$ limit 
\begin{eqnarray} 
\Gamma^{\wedge\wedge\vee}_{C1}&=&-\frac{g^3}{8\pi^2T_0}{K^\wedge\over M} 
\left\{\sum_{k=1}^\infty\sum_{l=1}^{M_1-1} 
{k\over l}\psi^\prime\left(1+k{M_1(M_2+l)\over lM}\right)\right. 
\nonumber\\ 
&&\hskip1in\left.\left[f_{k}{(2M_1-l)(2M_2+l)\over l^2}+h_{k}\right] 
+ \;(\; 1\; \leftrightarrow\; 2\;)\;\right\}\; . 
\label{c1triangle} 
\end{eqnarray} 
 
For the $C_2$ contribution, $k_2^\prime$ is limited by $f,h$. Then $k_1$ ranges 
freely from $1$ to $\infty$ for $l>0$, but for $l<0$, $1\leq k_1^\prime 
\leq k_2^\prime-1$. In the first case, only the second term of $H$ can get 
large, and in the second case no term can get large. So with 
$u_1=e^{-p_1^2/2M_1T_0}$, we need Eq.~\ref{c1sum} (with $u\to u_1$ 
and $k_2^\prime\to k_1^\prime$) for $l>0$ and 
\begin{eqnarray} 
\sum_{k^\prime_1=1}^{k_2^\prime-1} 
{1\over(k^\prime_1+z)^2}&=& \psi^\prime(z+1) -\psi^\prime(z+k_2^\prime +1) 
-{1\over (k_2^\prime+z)^2} 
\qquad{\rm for}\; l<0. 
\end{eqnarray} 
The $C_3$ contribution is similar but with the roles of $l>0$ 
and $l<0$ switched. In fact, inspection shows that interchanging 
$1\leftrightarrow2$ takes the $C_2$ contribution for $l>0$ ($l<0$) 
into the $C_3$ contribution for $l<0$ ($l>0$). So we need only 
display the $l>0$ cases explicitly. Combining all three contributions we 
have for $a\to0$ 
\begin{eqnarray} 
&&\hskip-.25in\Gamma^{\wedge\wedge\vee}_{C1} 
+\Gamma^{\wedge\wedge\vee}_{C2} 
+\Gamma^{\wedge\wedge\vee}_{C3}\approx-\frac{g^3}{8\pi^2T_0}{K^\wedge\over M} 
\left\{\sum_{k=1}^\infty\sum_{l=1}^{M_1-1} 
{k\over l}\psi^\prime\left(1+k{M_1(M_2+l)\over lM}\right) 
\right.\nonumber\\&&\hskip2.3in\times\left. 
\left[f_{k}{(2M_1-l)(2M_2+l)\over l^2}+h_{k}\right] 
\right.\nonumber\\  
&&\hskip-.2in+\left.\sum_{k=1}^\infty\sum_{l=1}^{M_1-1} 
{k(M_1-l)^2M\over (M_2+l)M_1^3}\psi^\prime\left(1+k{lM\over M_1(M_2+l)}\right) 
%\right.\nonumber\\ &&\hskip3.5in\times\left. 
\left[f_{k}{(M_2-l)(M+M_1-l)\over (M_2+l)^2}+h_{k}\right] 
\right.\nonumber\\  
&&\hskip-.2in+\left.\sum_{k=1}^\infty\sum_{l=1}^{M_1-1} 
{k(M_2+l)^2M\over (M_1-l)M_2^3} 
\left[{M_2^2(M_1-l)^2\over k^2M_1^2(M_2+l)^2}  
+ \psi^\prime\left(1+k{M_1(M_2+l)\over M_2(M_1-l)}\right) 
-\psi^\prime\left(1+k{lM\over M_2(M_1-l)}\right)\right] 
\right.\nonumber\\&&\hskip2.2in\times\left. 
\left[f_{k}{(M_1+l)(M+M_2+l)\over (M_1-l)^2}+h_{k}\right] 
%\right.\nonumber\\&&\left.  
+ \;(\; 1\; \leftrightarrow\; 2\;)\;\right\}\; . 
\label{c123triangle} 
\end{eqnarray} 
 
The $l$ sum can be rewritten and evaluated approximately and the 
details can be found in Appendix A.3. This yields the result of the continuum 
limit of the triangle with internal longitudinal gluons:  
 
\begin{eqnarray} 
&&\hskip-.25in\Gamma^{\wedge\wedge\vee}_{II} 
\approx-\frac{g^3}{8\pi^2T_0}{K^\wedge\over M} 
\left. \Bigg\{\sum_{k=1}^\infty \int_\epsilon^{1}d\xi 
{k\over \xi}\psi^\prime\left(1+k{M_2+\xi M_1)\over \xi M}\right) 
\left[f_{k}{(2-\xi)(2M_2+\xi M_1)\over \xi^2M_1}+h_{k}\right] 
\right.\nonumber\\  
&&\hskip-.2in+\left.\sum_{k=1}^\infty\int_0^1 d\xi 
{k(1-\xi)^2M\over M_2+\xi M_1} 
\psi^\prime\left(1+k{\xi M\over M_2+\xi M_1}\right) 
%\right.\nonumber\\ &&\hskip3.5in\times\left. 
\left[f_{k}{(M_2-\xi M_1)(M+M_1(1-\xi))\over (M_2+\xi M_1)^2}+h_{k}\right] 
\right.\nonumber\\  
&&\hskip-.2in+\left.\sum_{k=1}^\infty\int_0^{1-\epsilon}d\xi 
{k(M_2+\xi M_1)^2M\over (1-\xi)M_2^3} 
\left[{M_2^2(1-\xi)^2\over k^2(M_2+\xi M_1)^2}  
+ \psi^\prime\left(1+k{M_2+\xi M_1\over
M_2(1-\xi)}\right)\right. \right. \nonumber\\  
&&\hskip-.2in \left. \left.
-\psi^\prime\left(1+k{\xi M\over M_2(1-\xi)}\right)\right]  
\left[f_{k}{(1+\xi)(M+M_2+\xi M_1)\over M_1(1-\xi)^2}+h_{k}\right] 
-{2M^2\over\epsilon M_1M_2} 
%\right.\nonumber\\&&\left.  
+ \;(\; 1\; \leftrightarrow\; 2\;)\;\right. \Bigg\}\nonumber\\ 
&&\hskip2cm-\frac{g^3}{4\pi^2T_0}{K^\wedge\over M}\left[+{2M\pi^2\over3}-{M^2\over M_1M_2} 
(\ln\epsilon^2 M_1M_2 +2\gamma)\left(3-{\pi^2\over6}\right)\right]. 
\label{triangleII} 
\end{eqnarray} 
As $\epsilon\to0$ the r.h.s. approaches a finite result. Then the 
divergence as $M\to\infty$ is contained entirely in the last line. 
Thus we can write in the continuum limit 
\begin{eqnarray} 
\Gamma^{\wedge\wedge\vee}_{II} 
&=&\Gamma^{\wedge\wedge\vee}_{II,{\rm finite}} 
-\frac{g^3K^\wedge}{4\pi^2T_0} 
\left[{2\pi^2\over3}-{M\over M_1M_2} 
(\ln M_1M_2 +2\gamma)\left(3-{\pi^2\over6}\right)\right]. 
\end{eqnarray} 
Notice that these divergent terms cancel some of the 
divergent terms in $\Gamma^{\wedge\wedge\vee}_I$ leading to the complete 
vertex 
%\newpage 
\begin{eqnarray} 
&&\hskip-.2in\Gamma^{\wedge\wedge\vee}= 
\Gamma_{{\rm finite}}^{\wedge\wedge\vee} 
+\frac{g^3}{16\pi^2T_0}{(p_1^+p_2^\wedge-p_2^+p_1^\wedge) p^+\over p_1^+p_2^+} 
\left\{ 
%\right.\nonumber\\&&\hskip1in+\left. 
\ln{p^*\over p_1^*} 
\left[-4\ln {p^+\over p^+_1}+{22\over9}\right] 
 +\ln{p^*\over p_2^*} 
\left[-4\ln {p^+\over p^+_2}+{22\over9}\right]\right\}\nonumber\\ 
&&+\frac{g^3}{4\pi^2T_0}\; 
{(p_1^+p_2^\wedge-p_2^+p_1^\wedge) p^+\over p_1^+p_2^+} \left\{ 
{1\over2}\sum_{i=1}^3\ln{2 \over a|p_i^*|} 
\left[-4\left(\ln|M_i|+\gamma\right)+{22\over9}\right]\right.\nonumber\\ 
&&\hskip1.5in+(\ln M_1 
+\gamma)\left[\left({p^*+p_1^*\over p^*-p_1^*} 
-{p_1^*-p_2^*\over p_1^*+p_2^*}\right) 
\ln{p^*\over p_1^*}\right] \nonumber\\ 
&&\hskip1.5in\left. +(\ln M_2 
+\gamma)\left[\left({p^*+p_2^*\over p^*-p_2^*} 
+{p_1^*-p_2^*\over p_1^*+p_2^*}\right) 
\ln{p^*\over p_2^*}\right]  \right\}\; . 
\label{totaltriangle} 
\end{eqnarray} 
\noindent 
where for further brevity we have defined the continuum limit finite 
variables $p_i^*\equiv{P_i^*}/{m}={p_1^2}/{p_1^+}$. 
 
As already discussed, 
the uv divergences in the second line are canceled up 
to the standard asymptotic freedom result by the wave function 
renormalization factors. The remaining divergences in the last two lines 
are only linear in $\ln M_i$, and are unavoidable in the off-shell 
amplitude. Notice that there is a special off-shell point, 
$p_1^*=p_2^*=-p^*$, for which they disappear. The finite part 
depends in detail on the choice of $f_k$, $h_k$. The requirement 
of Lorentz invariance is expected to limit these parameters sharply, 
if not over-determine them. If the latter holds, additional counter-terms 
will be required to achieve Lorentz invariance. 
 
\section{Conclusions} 
 
In this article we have extended previous work \cite{beringrt} 
by further exploring the discretized $SU(N_c)$ gauge theory proposed 
there. Although the discretized theory is completely regulated,  
there is no guarantee that gauge invariance has been respected. 
Since we have chosen a non-covariant gauge, violations 
of gauge invariance would show up as violations of Lorentz 
invariance in the continuum limit. 
It is therefore important to check 
whether it reproduces known weak coupling results. We have shown here 
that to one loop order we obtain the correct renormalization  
of the coupling, Eq.~(\ref{charge}). The remaining infra-red 
divergences shown in the last two lines of Eq.~(\ref{totaltriangle}) 
must be considered in the context of a physical 
quantity. There is no reason {\it a priori} to expect these 
divergences to disappear until one considers fully onshell, color 
singlet external states. It is however reassuring that only 
single logarithmic divergences appear. As pointed out earlier these 
divergences do disappear at the special offshell point 
$p_1^2/p_1^+=p_2^2/p_2^+=p_3^2/p_3^+ \ne 0$. 
 
The simplest on shell scattering process is gluon-gluon 
scattering. Thus the complete resolution of the 
remaining infra-red divergence issues at one loop must 
await the analysis of one-loop four gluon amplitudes, 
the obvious next step in this investigation. 
As confidence is gained that the discretized theory is 
faithful to the gauge invariant continuum theory, 
the application of the formalism to calculate QCD fishnet diagrams or 
to formulate a worldsheet 
description of QCD in the spirit of \cite{bardakcit} 
becomes more compelling.  
%%%%%% 
%%%%%% 
\appendix 
\setcounter{equation}0 
\renewcommand{\theequation}{\thesection.\arabic{equation}} 
 
\section{Details of Calculations} 
 
\subsection{Evaluation of $\Gamma^{\wedge\wedge\vee}_{B1}$} 
 
In the calculation of $\Gamma^{\wedge\wedge\vee}_{B1}$ we start by integrating by parts. 
This transfers the derivative 
to the factor $(e^{-H}-e^{-H_0})$. For definiteness take the 
case $l>0$. Then we compute 
\begin{equation}{\partial\over\partial T_2}(e^{-H}-e^{-H_0}) 
=-e^{-H}{H-T_1lP_1^*\over T_2}+e^{-H_0}{H_0-T_1lP_1^*\over T_2} 
+e^{-H}{K^2\over M_1}\left[{(M_1-l)T_1T_2\over (MT_2+M_1T_1)^2}\right]. 
\end{equation} 
The first two terms on the r.h.s. partly cancel after integration over $T_1,T_2$. 
This is because the integrals are separately finite, so one can change  
variables $T_1=T_2T$ in each term separately. For the first term 
we find 
\begin{eqnarray} 
&&\hskip-.5in-\int_0^\infty dT_1dT_2 {\cal 
I}(T_1,T_2){H(T_1,T_2)-T_1lP_1^*\over T_2}  
e^{-H(T_1,T_2)}=\nonumber\\ 
&&\hskip-.25in-\int dTdT_2 {\cal I}(T,1) 
\left[H(T,1)-TlP_1^* \right]e^{-T_2H(T,1)} 
= -\int dT{\cal I}(T,1)\left[1-{TlP_1^*\over H(T,1)}\right], 
\end{eqnarray} 
and the second term yields the same expression with $H(T,1)\to H_0(T,1)$,  
so the two terms combine to 
\begin{eqnarray} 
\int_0^\infty dT{\cal I}(T,1)Tl{H_0(T,1)-H(T,1)\over H(T,1)(lT+M_2+l)}. 
\end{eqnarray}  
Simplifying the contribution to the $T$ integrand from these 
terms leads to the continuum limit 
\begin{eqnarray} 
\Gamma^{\wedge\wedge\vee}_{B1}&\to&{g^3K^\wedge\over4\pi^2 T_0}{M\over M_1M_2} 
\left\{\sum_{l=1}^{M_1-1}\int_0^\infty dT% 
\left[{lT(H_0-H)\over H(lT+M_2+l)}+{(M_1-l)TK^2\over 
HM_1(M+M_1T)^2}\right] 
{(M_1-l)^2\over(M+M_1T)^2}\right.\nonumber\\ 
&& 
\hskip-1cm\left.\left[{T\over2M}B_1+{M_1T+2M\over2M^2}B_2 
+{M_2T(M_1-l)+2MlT\over 2M^2(M_1-l)}B_3 
+{M_2+l\over M(M_1-l)}B_3\right]\; + \; (\;1\leftrightarrow\; 2\;)\right\}\; 
\nonumber\\ 
&=&{g^3K^\wedge\over4\pi^2 T_0}{M\over M_1M_2} 
\left\{\sum_{l=1}^{M_1-1}\int_0^\infty dT% 
\left[{lT(H_0-H)\over H(lT+M_2+l)}+{(M_1-l)TK^2\over 
HM_1(M+M_1T)^2}\right] 
{(M_1-l)^2\over(M+M_1T)^2}\right.\nonumber\\ 
&&\hskip1.5in\left. 
{M_1\over M^2}\left[{M+M_1T\over M_1-l}A^\prime-{lT+M_2+l\over M_1-l}A\right] 
\; + \; (\;1\leftrightarrow\; 2\;)\right\}\;, 
\end{eqnarray} 
where we have defined 
\begin{eqnarray} 
A^\prime &=& {M^2M_2^2\over lM_1(M_2+l)^2}-{M_2(M_2+l)^2\over M_1M(M_1-l)} 
-{M_2(M_1-l)^3\over M_1M(M_2+l)^2}. 
\end{eqnarray} 
 
Notice that this result combines neatly with $\Gamma^{\wedge\wedge\vee}_A$ to give: 
\begin{eqnarray} 
\Gamma^{\wedge\wedge\vee}_A+\Gamma^{\wedge\wedge\vee}_{B1}&\to&{g^3K^\wedge\over4\pi^2
T_0}{M\over M_1M_2}  
\left\{\sum_{l=1}^{M_1-1}\int_0^\infty dT% 
\left[{TK^2(M_1-l)^2A^\prime\over M^2H(M+M_1T)^3}
\right.\right. 
\nonumber\\ 
&&\hskip-2.2cm\left.\left.
+{lT(H_0-H)(M_1-l)M_1A^\prime 
\over M^2H(lT+M_2+l)(M+M_1T)} 
\mbox{}-{lT(H_0-H)(M_1-l)M_1A\over M^2H(M+M_1T)^2}\right] 
\; + \; (\;1\leftrightarrow\; 2\;)\right\}\,. 
\end{eqnarray} 
 
\subsection{Evaluation of $\Gamma^{\wedge\wedge\vee}_{B2}$} 
 
We analyze the continuum limit of the $\Gamma^{\wedge\wedge\vee}_{B2}$ contribution 
to the $B$ terms, which will be retained 
as discrete sums over the $k$'s. Again for definiteness we display  
the case $l>0$ in detail: 
\begin{align} 
\Gamma_{B2}^{\wedge\wedge\vee}&=\frac{g^3K^\wedge}{16\pi^2T_0^3}{M\over M_1M_2} 
\left\{\sum_{l=1}^{M_1-1}\frac{1}{|l|(M_2+l)(M_1-l)} 
\sum_{k_1,k^\prime_2}  
        \left[{\frac{T_1B_1+T_2B_2+T_3B_3}{(T_1+T_2+T_3)^3}}\right] e^{-H_0} 
        +  (\,1\leftrightarrow\, 2\,)\right\} \nonumber\\ 
&=\frac{g^3K^\wedge}{4\pi^2T_0}{M\over M_1M_2} 
\left\{ \sum_{l=1}^{M_1-1}\frac{M_1-l}{|l|(M_2+l)} 
\sum_{k_1,k^\prime_2}  
\left[{\frac{k_1N_1+k^\prime_2N_2}{(k_1\alpha 
+k^\prime_2\beta)^3}}\right]u_1^{k_1+k_2^\prime} 
\; + \; (\;1\leftrightarrow\; 2\;)\right\}\;. 
\label{b2def} 
\end{align} 
where 
\begin{eqnarray} 
\alpha&\equiv&{M_1\over l},\quad \beta\equiv {M\over M_2+l},\quad 
N_1\equiv{B_1(M_1-l)\over l}+B_3,\quad N_2\equiv{B_2(M_1-l)\over M_2+l}+B_3 
\nonumber\\ 
u_1&\equiv& e^{-p_1^2/2M_1T_0}=e^{-ap_1^2/2p^+_1}, 
\qquad u_2\equiv e^{-p_2^2/2M_2T_0}=e^{-ap_2^2/2p_2^+}, 
\end{eqnarray} 
where $u_2$ is to be used in the case $k_1,l<0$ instead of $u_1$. 
Clearly the continuum limit entails $u_1,u_2\to 1$, causing the 
$k$ sums to diverge logarithmically. To make this explicit, we first note 
the integral representation 
\begin{eqnarray} 
\sum_{k_1,k^\prime_2}{u_1^{k_1+k_2^\prime}\over (k_1\alpha+k^\prime_2\beta)^2} 
&=&\int_0^\infty tdt {u_1^2\over(e^{\alpha t}-u_1)(e^{\beta t}-u_1)}\\ 
&\approx&\int_\epsilon^\infty tdt {1\over(e^{\alpha t}-1)(e^{\beta t}-1)} 
+\int_0^\epsilon {tdt\over(1-u_1+\alpha t)(1-u_1+\beta t)}. 
\end{eqnarray} 
where the approximate form is valid for $1-u_1\ll\epsilon\ll\alpha,\beta$. 
Doing the integral in the second term leads to: 
\begin{eqnarray} 
&&\hskip-.5in\sum_{k_1,k^\prime_2}{u_1^{k_1+k_2^\prime}\over (k_1\alpha+k^\prime_2\beta)^2} 
\approx{1\over\alpha\beta}\left[\ln{2p_1^+\over ap_1^2}+{\beta\ln\alpha 
-\alpha\ln\beta\over \beta-\alpha} 
\right] 
%\nonumber\\&&\hskip1in 
+{1\over\alpha\beta}\ln\epsilon+\int_\epsilon^\infty tdt {1\over(e^{\alpha t}-1)(e^{\beta t}-1)}. 
\end{eqnarray} 
The sums we require can be obtained from this identity by differentiation 
with respect to $\alpha$ or $\beta$. To present the results it is convenient 
to define a function $f(x)$ by 
\begin{eqnarray} 
f(x)&\equiv&{1\over2}{1+x\over1-x}\ln x + \lim_{\epsilon\to0}\left[ 
\ln\epsilon+\int_\epsilon^\infty{tdt\over(e^{t\sqrt{x}}-1) 
(e^{t/\sqrt{x}}-1)}\right]\\ 
&=&{\ln x\over1-x}-x\int_0^\infty dt e^{-xt}{1-xt-e^{-xt}\over 
(1-e^{-xt})^2}\ln(1-e^{-t}) 
\end{eqnarray} 
where the second form is obtained by integration by parts. 
It is evident from the first form that $f(x)=f(1/x)$. Also 
one can easily calculate $f(1)=-\pi^2/6$. From the 
second form one easily sees that $f(x)\sim \ln x$ for 
$x\to0$, whence from the symmetry, $f(x)\sim -\ln  x$ 
for $x\to\infty$. Exploiting the function $f$ and its 
symmetries, we deduce 
\begin{eqnarray} 
\sum_{k_1,k^\prime_2}{u_1^{k_1+k_2^\prime}\over (k_1\alpha+k^\prime_2\beta)^2} 
&\approx&{1\over\alpha\beta}\left[\ln{2p_1^+\over ap_1^2} 
+f\left({\alpha\over\beta}\right)\right]\\ 
\sum_{k_1,k^\prime_2}{k_1u_1^{k_1+k_2^\prime}\over (k_1\alpha+k^\prime_2\beta)^2} 
&\approx&{1\over2\alpha^2\beta}\left[\ln{2p_1^+\over ap_1^2} 
+f\left({\alpha\over\beta}\right)\right]-{1\over2\alpha\beta^2} 
f^\prime\left({\alpha\over\beta}\right)\\ 
\sum_{k_1,k^\prime_2}{k^\prime_2u_1^{k_1+k_2^\prime}\over (k_1\alpha+k^\prime_2\beta)^2} 
&\approx&{1\over2\alpha\beta^2}\left[\ln{2p_1^+\over ap_1^2} 
+f\left({\alpha\over\beta}\right)\right]+{1\over\beta^3} 
f^\prime\left({\alpha\over\beta}\right)\nonumber\\ 
&\approx&{1\over2\alpha\beta^2}\left[\ln{2p_1^+\over ap_1^2} 
+f\left({\beta\over\alpha}\right)\right]-{1\over\alpha^2\beta} 
f^\prime\left({\beta\over\alpha}\right). 
\end{eqnarray} 
Inserting these results into Eq.~\ref{b2def} produces 
\begin{eqnarray} 
\Gamma_{B2}^{\wedge\wedge\vee}&=&\frac{g^3K^\wedge}{8\pi^2T_0}{M\over M_1M_2}  
\left\{\left[\sum_{l=1}^{M_1-1}{M_1-l\over MM_1}\left({N_1l\over M_1} 
+{N_2(M_2+l)\over M}\right)\left(\ln{2p_1^+\over ap_1^2} 
+f\left({\alpha\over\beta}\right)\right)\right.\right.\nonumber\\ 
&&\left.\left.-\sum_{l=1}^{M_1-1}{M_1-l\over MM_1}\left({N_1l\over M_1} 
-{N_2(M_2+l)\over M}\right) 
{\alpha\over\beta}f^\prime\left({\alpha\over\beta}\right)\right] 
+ \quad (\;1\; \leftrightarrow\; 2\;)\quad\right\} 
\nonumber\\ 
&=&-\frac{g^3K^\wedge}{4\pi^2T_0}{M\over M_1M_2}  
\left\{\sum_{l=1}^{M_1-1}\left[{B^\prime\over M} 
\left(\ln{2p_1^+\over ap_1^2} 
+f\left({\alpha\over\beta}\right)\right)+{AM_2(M_1-l)^2\over M^3M_1} 
{\alpha\over\beta}f^\prime\left({\alpha\over\beta}\right)\right]\right. 
\nonumber\\ 
&&\hskip3in\left.+ \quad (\;1\; \leftrightarrow\; 2\;) 
\phantom{{\alpha\over\beta}}\right\}, 
\end{eqnarray} 
where we have defined 
\begin{eqnarray} 
B^\prime= 
{(M_1-l)^3\over M^2(M_2+l)}+{MM_1\over l(M_1-l)} 
+{(M_2+l)^3\over M^2(M_1-l)}\; . 
\end{eqnarray} 
 
\subsection{Evaluation of $\Gamma^{\wedge\wedge\vee}_{II}$} 
For large $M_i$, the sum over $l$ can be approximated by an 
integral over $\xi=l/M_1$ from $\epsilon\ll1$ to $1-\epsilon$, 
plus sums for $1\leq l\leq\epsilon M_1$ and $M_1(1-\epsilon)\leq l 
\leq M_1-1$ which contain the divergences. These divergences are 
only present in the first sum on the r.h.s. of Eq.~\ref{c123triangle} 
for $l\ll M_1$ and in the last sum for $M_1-l\ll M_1$. The 
middle sum contains no divergence and can be replaced 
by an integral from $0$ to $1$ with no $\epsilon$ cutoff. 
To extract these divergent contributions, 
we can use the large argument expansion of $\psi^\prime$ 
\begin{eqnarray} 
\psi^\prime(z+1)\sim {1\over z}- {1\over2z^2} +O({1\over z^3}), 
\end{eqnarray} 
to isolate them. It is thus evident that their coefficients will 
be proportional to the moments $\sum f_k/k^n, \sum h_k/k^n$ 
for $k=0,1$, which are precisely the moments constrained 
by the requirement that the gluon remain massless at one loop. 
 
For the endpoint near $l=0$,  we put $z=kM_1(M_2+l)/lM$ 
and write 
\begin{eqnarray} 
{k\over l}\psi^\prime(z+1)\sim {M\over M_1(M_2+l)} 
-{l\over2k}{M^2\over M_1^2(M_2+l)^2} 
+\ldots , 
\end{eqnarray} 
so the summand for small $l$ becomes 
\begin{eqnarray} 
&&\hskip-.25in\sum_k \Bigg\{ f_k{M(2M_1-l)(2M_2+l)\over l^2M_1(M_2+l)} 
+{h_kM\over M_1(M_2+l)} 
-{f_k\over2k}{M^2(2M_1-l)(2M_2+l)\over lM_1^2(M_2+l)^2} 
-{h_k\over2k}{lM^2\over M_1^2(M_2+l)^2}\Bigg\} \nonumber\\ 
&&\hskip-.2in \sim {M(2M_1-l)(2M_2+l)\over l^2M_1(M_2+l)} 
+{M\over M_1(M_2+l)} 
-{\pi^2\over12}{M^2(2M_1-l)(2M_2+l)\over lM_1^2(M_2+l)^2} 
+{\pi^2(l-1) \over36l}{lM^2\over M_1^2(M_2+l)^2}\nonumber\\ 
&&\sim {4M\over l^2} 
-{2M^2\over lM_1M_2} 
-{\pi^2\over3}{M^2\over lM_1M_2}. 
\end{eqnarray} 
Summing $l$ up to $\epsilon M_1$ gives 
\begin{eqnarray} 
\sum_{l=1}^{\epsilon M_1}\left[ {4M\over l^2} 
-{2M^2\over lM_1M_2}-{\pi^2\over3}{M^2\over lM_1M_2}\right] 
\approx {4M\pi^2\over6}-{4M\over\epsilon M_1}-{M^2\over M_1M_2} 
(\ln\epsilon M_1 +\gamma)\left(2+{\pi^2\over3}\right), 
\end{eqnarray} 
Inserting these results into Eq.~\ref{c1triangle} and writing 
out explicitly the $1\leftrightarrow2$ terms for the divergent 
part gives 
\begin{eqnarray} 
\Gamma^{\wedge\wedge\vee}_{C1}&\approx&-\frac{g^3}{8\pi^2T_0}{K^\wedge\over M} 
\left\{\sum_{k=1}^\infty\int_{\epsilon}^{1}d\xi 
{k\over \xi}\psi^\prime\left(1+k{(M_2+\xi M_1)\over \xi M}\right) 
%\right.\nonumber\\&&\quad\left. 
\left[f_{k}{(2-\xi)(2M_2+\xi M_1)\over \xi^2 M_1}+h_{k}\right] 
\right.\nonumber\\&&\quad\left. 
+ \;(\; 1\; \leftrightarrow\; 2\;)\; 
-{4M^2\over\epsilon M_1M_2}+{8M\pi^2\over6}-{M^2\over M_1M_2} 
(\ln\epsilon^2 M_1M_2 +2\gamma)\left(2+{\pi^2\over3}\right)\right\}\; . 
\end{eqnarray} 
 
Only the third sum contributes near $l=M_1$. We again use the large argument 
expansion of $\psi^\prime$. But this time one only gets 
a logarithmic divergence, because the difference of $\psi^\prime$'s 
is of order $(M-l)^2$ as is the explicit rational term. Putting 
$z_1=kM_1(M_2+l)/M_2(M_1-l)$ and $z_2=klM/M_2(M_1-l)$, we have 
\begin{eqnarray} 
\psi^\prime(z_1+1)-\psi^\prime(z_2+1)&\sim&  
\left({1\over z_1}-{1\over z_2}\right)\left(1+{1\over 2z_1}+{1\over 2z_2}\right)\nonumber\\ 
&\sim&  -{M_2^2(M_1-l)^2\over klMM_1(M_2+l)}\left(1+O(M_1-l)\right)\sim 
 -{M_2^2(M_1-l)^2\over kM_1^2M^2}. 
\end{eqnarray} 
Thus the $l\sim M_1$ endpoint divergence is just  
\begin{eqnarray} 
-{g^3\over8\pi^2}{K^\wedge\over M}{4M^2\over M_1M_2} 
\sum_{k=1}^\infty \left({f_k\over k}-f_k\right) 
\sum_{l=M_1(1-\epsilon)}^{M_1-1}{1\over M_1-l} 
=-{g^3\over8\pi^2}{4MK^\wedge\over M_1M_2} 
\left({\pi^2\over6}-1\right) 
(\ln\epsilon M_1 +\gamma). 
\end{eqnarray} 
Putting everything together we obtain for the continuum 
limit of the triangle with internal longitudinal gluons 
\begin{eqnarray} 
&&\hskip-.25in\Gamma^{\wedge\wedge\vee}_{II} 
\approx-\frac{g^3}{8\pi^2T_0}{K^\wedge\over M} 
\left. \Bigg\{\sum_{k=1}^\infty \int_\epsilon^{1}d\xi 
{k\over \xi}\psi^\prime\left(1+k{M_2+\xi M_1)\over \xi M}\right) 
\left[f_{k}{(2-\xi)(2M_2+\xi M_1)\over \xi^2M_1}+h_{k}\right] 
\right.\nonumber\\  
&&\hskip-.2in+\left.\sum_{k=1}^\infty\int_0^1 d\xi 
{k(1-\xi)^2M\over M_2+\xi M_1} 
\psi^\prime\left(1+k{\xi M\over M_2+\xi M_1}\right) 
%\right.\nonumber\\ &&\hskip3.5in\times\left. 
\left[f_{k}{(M_2-\xi M_1)(M+M_1(1-\xi))\over (M_2+\xi M_1)^2}+h_{k}\right] 
\right.\nonumber\\  
&&\hskip-.2in+\left.\sum_{k=1}^\infty\int_0^{1-\epsilon}d\xi 
{k(M_2+\xi M_1)^2M\over (1-\xi)M_2^3} 
\left[{M_2^2(1-\xi)^2\over k^2(M_2+\xi M_1)^2}  
+ \psi^\prime\left(1+k{M_2+\xi M_1\over M_2(1-\xi)}\right)
\right.
\right.\nonumber\\
&& \left.\left. -\psi^\prime\left(1+k{\xi M\over
M_2(1-\xi)}\right)\right] 
\left[f_{k}{(1+\xi)(M+M_2+\xi M_1)\over M_1(1-\xi)^2}+h_{k}\right] 
-{2M^2\over\epsilon M_1M_2} 
%\right.\nonumber\\&&\left.  
+ \;(\; 1\; \leftrightarrow\; 2\;)\;\right. \Bigg\} \nonumber\\ 
&&\hskip2cm -\frac{g^3}{4\pi^2T_0}{K^\wedge\over M}\left[+{2M\pi^2\over3}-{M^2\over M_1M_2} 
(\ln\epsilon^2 M_1M_2 +2\gamma)\left(3-{\pi^2\over6}\right)\right]. 
\end{eqnarray}

\section{Divergent Parts of Integrals and Sums} 
\label{seca} 
The $T$ integral in Eq.~\ref{ab1} can be evaluated by expanding 
the integrand 
\begin{equation} 
I \equiv \left.{M_1TK^2(M_1-l)^2A^\prime\over M^2H(M+M_1T)^3} 
+{lT(H_0-H)(M_1-l)M_1^2A^\prime 
\over M^2H(lT+M_2+l)(M+M_1T)} 
%\right.\nonumber\\&&\hskip1.5in 
\mbox{}-{lT(H_0-H)(M_1-l)M_1^2A\over M^2H(M+M_1T)^2}\right.\; . 
\end{equation} 
in partial fractions. First note that since $(M_1T+M)H$ is 
a quadratic polynomial, it may be factored as 
\begin{eqnarray} 
(M_1T+M)H=lp_1^2(T-T_+)(T-T_-) 
\end{eqnarray} 
where $T_-\sim -(K^2+M_1M_2p^2)/lMp_1^2$ and $T_+\sim 
-MM_2p^2/(K^2+M_1M_2p^2)$ when $l\ll M_1$.  
Then the partial fraction expansion reads 
\begin{eqnarray} 
I={R_1\over T-T_+}+{R_2\over T-T_-}+{R_3\over lT+M_2+l} 
+{R_4\over( M_1T+M)^2}+{R_5\over M_1T+M}, 
\end{eqnarray} 
with the $R_i$ independent of $T$. Of course the $R_i$ are such that 
$I$ falls off at least as $1/T^2$ for large $T$, \ie 
$R_1+R_2+R_3/l+R_5/M_1=0$. This identity is helpful for determining 
$R_5$. Thus we have 
\begin{eqnarray} 
\int_0^\infty dTI=-{R_1\ln(-T_+)}-{R_2\ln(-T_-)}+{R_3\over l}\ln{l\over M_2+l} 
+{R_4\over MM_1}+{R_5\over M_1}\ln{M_1\over M}. 
\end{eqnarray} 
The $R_i$ are given explicitly by 
\begin{eqnarray} 
R_1 &=& {M_1(M_1-l)T_+\over lM^2p_1^2(T_+-T_-)}\left[{(M_1-l)K^2A^\prime 
\over (M_1T_++M)^2}+{lp_1^2A^\prime} 
-{lp_1^2(M_2+l+lT_+)A\over M_1T_++M }\right]\\ 
R_2 &=& -{M_1(M_1-l)T_-\over lM^2p_1^2(T_+-T_-)}\left[{(M_1-l)K^2A^\prime 
\over(M_1T_-+M)^2}+lp_1^2A^\prime-{lp_1^2(M_2+l+lT_-)A\over M_1T_-+M }\right]\\ 
R_3 &=& -l{(M_2+l)M_1^2A^\prime\over M^2M_2}\\ 
R_4 &=& M_1^2(M_1-l){A^\prime\over M}-{lM_1(M_1-l){A\over M}}\\ 
R_5 &=& {lM_1(M_1-l)A\over M^2}\left(1-{p_1^2MM_2\over K^2}\right) 
+{lM_1^2A^\prime\over MM_2}\nonumber\\ 
&&-{M_1^2(M_1-l)A^\prime\over M^2} 
\left[1-{M\over M_1T_++M}-{M\over M_1T_-+M }\right]. 
\end{eqnarray} 
When the $M$'s are large, the sum over $l$ can be replaced by 
an integral over $\xi=l/M_1$ as long as $\xi$ is kept away from 
the endpoints $\xi=0,1$. We can isolate the terms that 
give rise to singular end point contributions and simplify them considerably. 
We shall then separate the divergent contributions and 
display them in detail. 
 
First note that the worst endpoint divergence is $\xi^{-1}\ln\xi$ 
near $\xi=0$ or $1/(1-\xi)$ near $\xi=1$. Thus we can drop all 
terms down by a factor of $l/M_i$ for small $l$ or by $(M_1-l)/M_i$ 
for $l$ near $M_1$. Thus for $l\ll M_i$, we note that 
$lT_-(M+M_1T_+)\sim-K^2/p_1^2$ and $T_+/(M+M_1T_+)\sim-M_2p^2/K^2$ and obtain 
\begin{eqnarray} 
R_1 &\sim& -{p^2M_1^2M_2 
\over lK^2}{K^2+M_1M_2p^2-2M_2Mp_1^2\over (K^2+M_1M_2p^2) }= 
-{p^2M_1^2M_2\over lK^2}{M_1p_2^2-M_2p_1^2\over M_1p_2^2+M_2p_1^2 }\\ 
R_2 &\sim& -{M_1\over l}+{2M_1M_2p_1^2\over l(M_1p_2^2+M_2p_1^2)} 
=-{M_1\over l}{M_1p_2^2-M_2p_1^2\over M_1p_2^2+M_2p_1^2} 
\\ 
R_3 &\sim& -{M_1}\\ 
R_4 &\sim& -{MM_1^2\over l}\\ 
R_5 &\sim& {M_1^2\over l} 
\left[1+{MM_1p_2^2-MM_2p_1^2\over K^2}\right]. 
\end{eqnarray} 
Combining the $l\approx0$ endpoint contributions gives 
\begin{eqnarray} 
\int_0^\infty dTI&\sim&-R_1\ln{MM_2p^2\over K^2+M_1M_2p^2}-R_2 
\ln{K^2+M_1M_2p^2\over lMp_1^2} 
+{R_3\over l}\ln{l\over M_2} 
+{R_4\over MM_1}+{R_5\over M_1}\ln{M_1\over M}\nonumber\\ 
&\sim& {M_1\over l}\left\{{M_1p_2^2-M_2p_1^2\over M_1p_2^2+M_2p_1^2 }\left[ 
{M_1M_2p^2\over K^2}\ln{MM_2p^2\over K^2+M_1M_2p^2} 
+\ln{K^2+M_1M_2p^2\over lMp_1^2}\right]\right.\nonumber\\ 
&&\hskip.75in\left. 
-\ln{l\over M_2}+\left[1+{MM_1p_2^2-MM_2p_1^2\over K^2}\right]\ln{M_1\over M} 
-{1}\right\}\nonumber\\ 
&\sim& {M_1\over l}\left\{M{M_1p_2^2-M_2p_1^2\over K^2}\ln{M_1M_2p^2\over K^2+M_1M_2p^2} 
\right.\nonumber\\&&\hskip.25in\left. 
+{M_1p_2^2-M_2p_1^2\over M_1p_2^2+M_2p_1^2 }\ln{(K^2+M_1M_2p^2)^2 
\over lM^2M_2p^2p_1^2} 
-\ln{lM\over M_1M_2}-{1}\right\}\;, \quad{\rm for}\; l\ll M_i\; . 
\end{eqnarray} 
For the other endpoint, $M_1-l\ll M_i$, the roots of the polynomial 
$(M+M_1T)H$ approach $T_1=-M/M_1$ and $T_2=-p^2/p_1^2$. Which of these 
roots is approached by $T_\pm$ depends on the parameter values, 
but since the formulae are symmetric under their interchange, we 
can choose to use the first in place of $T_+$ and the second in place of  
$T_-$. Since the denominator $M+M_1T_1=0$ in this limit, \ 
we need to carefully evaluate 
$${M_1-l\over M+M_1T_1}\sim-{M_1(Mp_1^2-M_1p^2)\over K^2}.$$ 
Then we obtain for small $M_1-l$, 
\begin{eqnarray} 
R_1 &\sim& {M_1M_2(Mp_1^2+M_1p^2)\over (M_1-l) K^2} 
-{2MM_1p_1^2\over(M_1-l)(Mp_1^2-M_1p^2)}\\ 
R_2 &\sim& -{2p^2M_1^2\over (M-l)(M_1p^2-Mp_1^2)} 
\\ 
R_3 &\sim& l{M_1\over M_1-l}\\ 
R_4 &\sim& -{2MM_1^2\over M_1-l}\\ 
R_5 &\sim& {M_1^2\over (M_1-l)}-{M_1^2M_2(Mp_1^2+M_1p^2)\over (M_1-l)K^2}. 
\end{eqnarray} 
Combining the $l\approx M_1$ endpoint contributions gives 
\begin{eqnarray} 
\int_0^\infty dTI&\sim&\left[R_1+{R_3+R_5\over M_1}\right] 
\ln{M_1\over M}-R_2\ln{p^2\over p_1^2} 
+{R_4\over MM_1}\;, \hskip1cm {\rm for}\; M_1-l\ll M_i\hskip.75in\nonumber\\ 
&\sim& {2M_1\over (M_1-l)}\left[ 
{M_1p^2\over M_1p^2-Mp_1^2}\ln{M_1p^2\over Mp_1^2}-1\right]. 
\end{eqnarray} 
 
In writing the $l$ sum as an integral these endpoint divergences can be 
separated by picking $\epsilon\ll1$ and summing $l$ in the ranges 
$1\leq l\leq \epsilon M_1$ and $M_1(1-\epsilon)\leq l\leq M_1-1$.  
For these parts of the sum the above approximations can be made and the 
sum evaluated: 
\begin{eqnarray} 
\sum_{l=1}^{\epsilon M_1}{1\over l}&=&\sum_{l=M_1(1-\epsilon)}^{M_1-1} 
{1\over M_1-l}=\psi(1+\epsilon M_1)+\gamma\sim\ln\epsilon M_1 
+\gamma\nonumber\\ 
\sum_{l=1}^{\epsilon M_1}{\ln l\over l}&=&-\gamma(\psi(1+\epsilon M)+\gamma) 
-\int_0^\infty dt\ln t{e^{-t}-e^{-M_1\epsilon t}\over 1-e^{-t}}\nonumber\\ 
&\sim& {1\over2}\ln^2(M_1\epsilon)+{\zeta(2)-\gamma^2\over2} 
+{1\over2}\int_0^\infty dt{t\ln^2 t\over e^t-1}\equiv 
{1\over2}\ln^2(M_1\epsilon)+C 
\end{eqnarray} 
 The rest of the sum is replaced by 
an integral over $\epsilon\leq \xi\leq1-\epsilon$ 
\begin{eqnarray} 
{1\over M_1}\sum_{l=1}^{M_1-1}\int_0^\infty IdT&\sim&  
\int_\epsilon^{1-\epsilon}d\xi  
\int_0^\infty IdT\nonumber\\ 
&& +(\ln\epsilon M_1 
+\gamma)\left[\left({2M_1p^2\over M_1p^2-Mp_1^2} 
+{M_1p_2^2-M_2p_1^2\over M_1p_2^2+M_2p_1^2 }\right) 
\ln{M_1p^2\over Mp_1^2}-3  \right.\nonumber\\ 
&&\left.\hskip.5in +\left(M{M_1p_2^2-M_2p_1^2\over K^2} 
-2{M_1p_2^2-M_2p_1^2\over M_1p_2^2+M_2p_1^2 }\right)\ln{M_1M_2p^2\over K^2+M_1M_2p^2} 
\right]\nonumber\\&& 
-\left[\ln{M\over M_1M_2}(\ln \epsilon M_1+\gamma) 
+{1\over2}\ln^2(\epsilon M_1)+C\right]{2M_1p_2^2\over M_1p_2^2+M_2p_1^2 } 
\label{integral} 
\end{eqnarray} 
Finally, we must extract the divergent contributions that 
arise from replacing the sums  
\begin{eqnarray} 
\sum_{l=1}^{M_1-1}S_l=- \sum_{l=1}^{M_1-1}\left[{M_1B^\prime\over M} 
f\left({\alpha\over\beta}\right) + {AM_2(M_1-l)^2\over M^3} 
{\alpha\over\beta}f^\prime\left({\alpha\over\beta}\right)\right] 
\end{eqnarray} 
in Eq.~\ref{finaltriangle} by an integral. First, for $l\approx M_1$, 
$\alpha/\beta\approx1$ and only the first term gives a singular 
endpoint contribution, 
\begin{eqnarray} 
\sum_{l=M_1(1-\epsilon)}^{M_1-1}S_l\sim 
-{2M_1f(1)}[\psi(1+\epsilon M_1)+\gamma] 
\sim{\pi^2\over3}{M_1}[\ln\epsilon M_1+\gamma]. 
\label{sum1} 
\end{eqnarray} 
On the other hand, for $l\approx0$, we have 
\begin{eqnarray} 
f\left({M_1M_2\over lM}\right)&\sim& -\ln{M_1M_2\over lM}-\int_0^\infty 
dt e^{-t}\ln t {1-t-e^{-t}\over(1-e^{-t})^2}=\ln{lM\over M_1M_2}\\ 
{M_1M_2\over lM}f^\prime\left({M_1M_2\over lM}\right)&\sim& 
-1+{ lM\over M_1M_2}\left[{\pi^2\over12}+\ln{M_1M_2\over lM} -1\right].  
\end{eqnarray} 
The integral in the first line is zero because the integrand is a  
derivative of a function vanishing at the endpoints. 
Inserting these approximations, we obtain 
\begin{eqnarray} 
\sum_{l=1}^{\epsilon M_1}S_l 
&\sim&{M_1^2M_2\over M}\sum_{l=1}^{\epsilon M_1}{2\over l^2}+ 
{M_1}\left\{\sum_{l=1}^{\epsilon M_1}{1\over l}\ln{lM\over M_1M_2} 
-[\psi(1+\epsilon M_1)+\gamma]{\pi^2\over6}\right\} 
\label{sum2} 
\end{eqnarray} 
Putting Eqs.\ref{integral},\ref{sum1},\ref{sum2} together, some simplification 
occurs and we obtain 
\begin{eqnarray} 
{1\over M_1}\sum_{l=1}^{M_1-1}\left[S(l/M_1) + \int_0^\infty IdT\right] 
&\sim& \int_\epsilon^{1-\epsilon}d\xi  
\int_0^\infty IdT+\int_\epsilon^{1-\epsilon}d\xi S(\xi) 
\nonumber\\ 
&& \hskip-1.5in\mbox{} +{M_1M_2\pi^2\over3M}-{2M_2\over\epsilon M}  
+ (\ln\epsilon M_1 
+\gamma)\left[\left({2M_1p^2\over M_1p^2-Mp_1^2} 
+{M_1p_2^2-M_2p_1^2\over M_1p_2^2+M_2p_1^2 }\right) 
\ln{M_1p^2\over Mp_1^2}  \right.\nonumber\\ 
&&\left.\hskip-1.5in +\left(M{M_1p_2^2-M_2p_1^2\over K^2} 
-2{M_1p_2^2-M_2p_1^2\over M_1p_2^2+M_2p_1^2 }\right)\ln{M_1M_2p^2\over K^2+M_1M_2p^2} 
-3+{\pi^2\over6}\right]\nonumber\\&&\hskip-1.5in 
-\left[\ln{M\over M_1M_2}(\ln \epsilon M_1+\gamma) 
+{1\over2}\ln^2(\epsilon M_1)+C\right] 
{M_1p_2^2-M_2p_1^2\over M_1p_2^2+M_2p_1^2 }\; . 
\label{intsum1sum2} 
\end{eqnarray} 
When we add the contribution with $1\leftrightarrow2$, the antisymmetry 
of some of the coefficients leads to further simplification as 
well as a reduction in the degree of divergence of some of the terms: 
%\newpage 
\begin{eqnarray} 
&&\hskip-.2in  {1\over M_1}\sum_{l=1}^{M_1-1}\left[S(l/M_1)  
+ \int_0^\infty IdT\right] 
+ \;(\;1\;\leftrightarrow\;2\;)\sim  
\nonumber\\ 
&&\int_\epsilon^{1-\epsilon}d\xi  
\int_0^\infty IdT+\int_\epsilon^{1-\epsilon}d\xi S(\xi) 
+ \;(\;1\;\leftrightarrow\;2\;) 
%\nonumber\\&& \mbox{}  
-{2\over\epsilon}  
-\ln\epsilon\; 
{p^+_1p_2^2-p^+_2p_1^2\over p^+_1p_2^2+p^+_2p_1^2 }\ln{p^+_1\over p^+_2} 
\nonumber\\ 
&&+\ln\epsilon\left[{2p^+_1p^2\over p^+_1p^2-p^+p_1^2} 
\ln{p^+_1p^2\over p^+p_1^2} 
+{p^+_1p_2^2-p^+_2p_1^2\over p^+_1p_2^2+p^+_2p_1^2 } 
\ln{p^+_1p_2^2\over p^+_2p_1^2}  
%\right.\nonumber\\&&\left.  
+{2p^+_2p^2\over p^+_2p^2-p^+p_2^2} 
\ln{p^+_2p^2\over p^+p_2^2}-6+{\pi^2\over3}  \right]\nonumber\\ 
&&+\ln{p^+_1\over p^+_2}\left[\left(M{M_1p_2^2-M_2p_1^2\over K^2} 
-2{p^+_1p_2^2-p^+_2p_1^2\over p^+_1p_2^2+p^+_2p_1^2 }\right) 
\ln{M_1M_2p^2\over K^2+M_1M_2p^2} 
%\right.\nonumber\\&&\left. 
+{1\over2}\ln{p^+_1p^+_2\over M^2}\; 
{p^+_1p_2^2-p^+_2p_1^2\over p^+_1p_2^2+p^+_2p_1^2 }\right]\nonumber\\ 
&&+{2\pi^2M_1M_2\over3M}+ \left[(\ln M_1 
+\gamma)\left\{\left({2M_1p^2\over M_1p^2-Mp_1^2} 
+{M_1p_2^2-M_2p_1^2\over M_1p_2^2+M_2p_1^2 }\right) 
\ln{M_1p^2\over Mp_1^2}-3+{\pi^2\over6} \right\} \right.\nonumber\\ 
&&\left. +(\ln M_2 
+\gamma)\left\{\left({2M_2p^2\over M_2p^2-Mp_2^2} 
-{M_1p_2^2-M_2p_1^2\over M_1p_2^2+M_2p_1^2 }\right) 
\ln{M_2p^2\over Mp_2^2}-3+{\pi^2\over6} \right\}  \right]\; . 
\label{12intsum1sum2} 
\end{eqnarray} 
As $\epsilon\to0$ the first two lines on the r.h.s. approach a finite 
$\epsilon$ independent answer. The third line is explicitly finite.  
All divergences are shown in the last two lines. As $M_i\to\infty$ 
there is a leading linear divergence as well as a single 
logarithmic sub-leading divergence.  

\underline{Acknowledgments:}  
This work was supported in  
part by the Department of Energy under Grant No. DE-FG02-97ER-41029.

%\bibliography{../../sheet/larefs} 
%\bibliographystyle{unsrt} 
\end{document}